\documentclass[aps,prl,reprint,amsmath,amssymb,longbibliography]{revtex4-2}
\ifdefined\pdfglyphtounicode
  \pdfglyphtounicode{parenleftbig}{0028 FE01}
  \pdfglyphtounicode{parenrightbig}{0029 FE01}
  \pdfglyphtounicode{bracketleftbig}{005B FE01}
  \pdfglyphtounicode{bracketrightbig}{005D FE01}
  \pdfglyphtounicode{floorleftbig}{230A FE01}
  \pdfglyphtounicode{floorrightbig}{230B FE01}
  \pdfglyphtounicode{ceilingleftbig}{2308 FE01}
  \pdfglyphtounicode{ceilingrightbig}{2309 FE01}
  \pdfglyphtounicode{braceleftbig}{007B FE01}
  \pdfglyphtounicode{bracerightbig}{007D FE01}
  \pdfglyphtounicode{angbracketleftbig}{27E8 FE01}
  \pdfglyphtounicode{angbracketrightbig}{27E9 FE01}
  \pdfglyphtounicode{vextendsingle}{20D3}
  \pdfglyphtounicode{vextenddouble}{20E6}
  \pdfglyphtounicode{slashbig}{002F FE01}
  \pdfglyphtounicode{backslashbig}{005C FE01}
  \pdfglyphtounicode{parenleftBig}{0028 FE02}
  \pdfglyphtounicode{parenrightBig}{0029 FE02}
  \pdfglyphtounicode{parenleftbigg}{0028 FE03}
  \pdfglyphtounicode{parenrightbigg}{0029 FE03}
  \pdfglyphtounicode{bracketleftbigg}{005B FE03}
  \pdfglyphtounicode{bracketrightbigg}{005D FE03}
  \pdfglyphtounicode{floorleftbigg}{230A FE03}
  \pdfglyphtounicode{floorrightbigg}{230B FE03}
  \pdfglyphtounicode{ceilingleftbigg}{2308 FE03}
  \pdfglyphtounicode{ceilingrightbigg}{2309 FE03}
  \pdfglyphtounicode{braceleftbigg}{007B FE03}
  \pdfglyphtounicode{bracerightbigg}{007D FE03}
  \pdfglyphtounicode{angbracketleftbigg}{27E8 FE03}
  \pdfglyphtounicode{angbracketrightbigg}{27E9 FE03}
  \pdfglyphtounicode{slashbigg}{002F FE03}
  \pdfglyphtounicode{backslashbigg}{005C FE03}
  \pdfglyphtounicode{parenleftBigg}{0028 FE04}
  \pdfglyphtounicode{parenrightBigg}{0029 FE04}
  \pdfglyphtounicode{bracketleftBigg}{005B FE04}
  \pdfglyphtounicode{bracketrightBigg}{005D FE04}
  \pdfglyphtounicode{floorleftBigg}{230A FE04}
  \pdfglyphtounicode{floorrightBigg}{230B FE04}
  \pdfglyphtounicode{ceilingleftBigg}{2308 FE04}
  \pdfglyphtounicode{ceilingrightBigg}{2309 FE04}
  \pdfglyphtounicode{braceleftBigg}{007B FE04}
  \pdfglyphtounicode{bracerightBigg}{007D FE04}
  \pdfglyphtounicode{angbracketleftBigg}{27E8 FE04}
  \pdfglyphtounicode{angbracketrightBigg}{27E9 FE04}
  \pdfglyphtounicode{slashBigg}{002F FE04}
  \pdfglyphtounicode{backslashBigg}{005C FE04}
  \pdfglyphtounicode{slashBig}{002F FE02}
  \pdfglyphtounicode{backslashBig}{005C FE02}
  \pdfglyphtounicode{parenlefttp}{239B}
  \pdfglyphtounicode{parenrighttp}{239E}
  \pdfglyphtounicode{bracketlefttp}{23A1}
  \pdfglyphtounicode{bracketrighttp}{23A4}
  \pdfglyphtounicode{bracketleftbt}{23A3}
  \pdfglyphtounicode{bracketrightbt}{23A6}
  \pdfglyphtounicode{bracketleftex}{23A2}
  \pdfglyphtounicode{bracketrightex}{23A5}
  \pdfglyphtounicode{bracelefttp}{23A7}
  \pdfglyphtounicode{bracerighttp}{23AB}
  \pdfglyphtounicode{braceleftbt}{23A9}
  \pdfglyphtounicode{bracerightbt}{23AD}
  \pdfglyphtounicode{braceleftmid}{23A8}
  \pdfglyphtounicode{bracerightmid}{23AC}
  \pdfglyphtounicode{braceex}{23AA}
  \pdfglyphtounicode{arrowvertex}{23D0}
  \pdfglyphtounicode{parenleftbt}{239D}
  \pdfglyphtounicode{parenrightbt}{23A0}
  \pdfglyphtounicode{parenleftex}{239C}
  \pdfglyphtounicode{parenrightex}{239F}
  \pdfglyphtounicode{angbracketleftBig}{27E8 FE02}
  \pdfglyphtounicode{angbracketrightBig}{27E9 FE02}
  \pdfglyphtounicode{unionsqtext}{2A06 FE01}
  \pdfglyphtounicode{unionsqdisplay}{2A06 FE02}
  \pdfglyphtounicode{contintegraltext}{222E FE01}
  \pdfglyphtounicode{contintegraldisplay}{222E FE02}
  \pdfglyphtounicode{circledottext}{2A00 FE01}
  \pdfglyphtounicode{circledotdisplay}{2A00 FE02}
  \pdfglyphtounicode{circleplustext}{2A01 FE01}
  \pdfglyphtounicode{circleplusdisplay}{2A01 FE02}
  \pdfglyphtounicode{circlemultiplytext}{2A02 FE01}
  \pdfglyphtounicode{circlemultiplydisplay}{2A02 FE02}
  \pdfglyphtounicode{summationtext}{2211 FE01}
  \pdfglyphtounicode{producttext}{220F FE01}
  \pdfglyphtounicode{integraltext}{222B FE01}
  \pdfglyphtounicode{uniontext}{22C3 FE01}
  \pdfglyphtounicode{intersectiontext}{22C2 FE01}
  \pdfglyphtounicode{unionmultitext}{2A04 FE01}
  \pdfglyphtounicode{logicalandtext}{22C0 FE01}
  \pdfglyphtounicode{logicalortext}{22C1 FE01}
  \pdfglyphtounicode{summationdisplay}{2211 FE02}
  \pdfglyphtounicode{productdisplay}{220F FE02}
  \pdfglyphtounicode{integraldisplay}{222B FE02}
  \pdfglyphtounicode{uniondisplay}{22C3 FE02}
  \pdfglyphtounicode{intersectiondisplay}{22C2 FE02}
  \pdfglyphtounicode{unionmultidisplay}{2A04 FE02}
  \pdfglyphtounicode{logicalanddisplay}{22C0 FE02}
  \pdfglyphtounicode{logicalordisplay}{22C1 FE02}
  \pdfglyphtounicode{coproducttext}{2210 FE01}
  \pdfglyphtounicode{coproductdisplay}{2210 FE02}
  \pdfglyphtounicode{hatwide}{02C6 FE01}
  \pdfglyphtounicode{hatwider}{02C6 FE02}
  \pdfglyphtounicode{hatwidest}{02C6 FE03}
  \pdfglyphtounicode{tildewide}{02DC FE01}
  \pdfglyphtounicode{tildewider}{02DC FE02}
  \pdfglyphtounicode{tildewidest}{02DC FE03}
  \pdfglyphtounicode{bracketleftBig}{005B FE02}
  \pdfglyphtounicode{bracketrightBig}{005D FE02}
  \pdfglyphtounicode{floorleftBig}{230A FE02}
  \pdfglyphtounicode{floorrightBig}{230B FE02}
  \pdfglyphtounicode{ceilingleftBig}{2308 FE02}
  \pdfglyphtounicode{ceilingrightBig}{2309 FE02}
  \pdfglyphtounicode{braceleftBig}{007B FE02}
  \pdfglyphtounicode{bracerightBig}{007D FE02}
  \pdfglyphtounicode{radicalbig}{221A FE01}
  \pdfglyphtounicode{radicalBig}{221A FE02}
  \pdfglyphtounicode{radicalbigg}{221A FE03}
  \pdfglyphtounicode{radicalBigg}{221A FE04}
  \pdfglyphtounicode{radicalbt}{23B7}
  \pdfglyphtounicode{radicalvertex}{20D3}
  \pdfglyphtounicode{radicaltp}{231C}
  \pdfglyphtounicode{arrowvertexdbl}{20E6}
  \pdfglyphtounicode{arrowtp}{2191}
  \pdfglyphtounicode{arrowbt}{2193}
  \pdfglyphtounicode{bracehtipdownleft}{23DF}
  \pdfglyphtounicode{bracehtipdownright}{23DF}
  \pdfglyphtounicode{bracehtipupleft}{23DE}
  \pdfglyphtounicode{bracehtipupright}{23DE}
  \pdfglyphtounicode{arrowdbltp}{21D1}
  \pdfglyphtounicode{arrowdblbt}{21D3}
\fi
\usepackage[T1]{fontenc}
\usepackage{lmodern,graphicx,bm,mathtools,microtype,xcolor}
\definecolor{linkblue}{HTML}{173D56}
\usepackage[colorlinks=true,allcolors=linkblue]{hyperref}
\hypersetup{pdftitle={Universal Critical Jump for Inverse-Square Ising Ferromagnets},pdfauthor={Dyuman Bhattacharya}}
\newcommand{\E}{\mathbb E}
\newcommand{\ind}{\mathbf1}

\newcommand{\R}{\mathbb R}

\begin{document}
\title{Universal Critical Jump for Inverse-Square Ising Ferromagnets}
\author{Dyuman Bhattacharya}
\email{d7bhatta@uwaterloo.ca}
\affiliation{Department of Physics and Astronomy, University of Waterloo, Waterloo, Ontario, Canada}
\date{October 26, 2026}
\begin{abstract}
For every translation-invariant ferromagnetic Ising chain with
$n^2J_n\to B>0$, we prove
$2\beta_cBm_c^2=1$, without restrictions on the finite core or the
tail convergence rate. The proof bounds the energy of macroscopic
interfaces and rules them out if the critical product exceeds one.
Despite the nonzero jump, connected susceptibility is infinite both
at criticality and in a neighboring ordered interval. At the critical
coupling, the excess magnetization decays more slowly than any
positive power of the field as the field tends to zero.
We also prove the jump relation in
continuous time, for a positive-Stieltjes class of Ohmic spin-boson models.
\end{abstract}
\maketitle

Long-range order in the inverse-square Ising chain begins with a jump.
The magnetization is
nonzero at the ordering threshold, the Thouless
effect~\cite{Thouless,ACCN}, while correlations can decay slowly
inside the ordered phase~\cite{IN}. An inverse-square interaction
also arises in imaginary time when an Ohmic oscillator bath is
integrated out of a spin-boson model~\cite{AYH,Leggett,Spohn}.
Determining the jump is therefore also a question about localization
of a dissipative quantum spin.

For an interaction $J_n\sim B/n^2$, the established ordered-phase
inequality is $2\beta Bm(\beta)^2\ge1$~\cite{ACCN,DGT}.
Imbrie and Newman asked whether this bound becomes an equality at
criticality for each fixed microscopic interaction, and described
the consequences for critical correlations~\cite[Sec.~1(iv)]{IN}.
The equality agrees with renormalization-group predictions and
numerics~\cite{AYH,LuijtenMessingfeld}; finite-size extrapolation
is delicate because convergence is slow~\cite{Tower2026}.
We prove the equality under the sole tail assumption $n^2J_n\to B>0$.
To recover the magnetization in its coefficient, we ask how much
pressure is lost when a small amount of interaction is removed at
large separations.

\emph{Unrestricted critical jump.---}
For spins $\sigma_i=\pm1$ on $\mathbb Z$, use the Gibbs exponent
\begin{equation}\label{main:weight}
 \beta\sum_{i<j}J_{j-i}\sigma_i\sigma_j+h\sum_i\sigma_i.
\end{equation}
Each unordered pair is counted once, and $h$ is the dimensionless
field. Let $p_\beta(h)$ be the pressure,
$m(\beta)=\partial_h^+p_\beta(0)$, and
$\beta_c=\inf\{\beta:m(\beta)>0\}$. Assume only
\begin{equation}\label{main:tail}
 J_n\in[0,\infty),\qquad n^2J_n\longrightarrow B\in(0,\infty)
\end{equation}
Then $0<\beta_c<\infty$, $m_c:=m(\beta_c)>0$, and
\begin{equation}\label{main:jump}
 \boxed{\,2\beta_cB\,m_c^2=1\,}.
\end{equation}
In particular, one may set any finite number of the $J_n$ to zero.
One may also let the tail oscillate; for example,
$J_n=B n^{-2}[1+(-1)^n/(2\log(n+2))]$.
This example has infinitely many increases $J_{n+1}>J_n$ and a
divergent remainder $\sum_n n|J_n-B/n^2|$.
Equation~\eqref{main:jump} still applies to this interaction, and
to any finite change of its couplings that preserves ferromagnetism.
It determines $m_c$ from $\beta_c$ and $B$, although neither
$\beta_c$ nor $m_c$ need be the same for different finite cores.
Proof details are in the Supplemental Material (SM)~\cite{SM}.

\begin{figure}[t]
\includegraphics[width=\columnwidth]{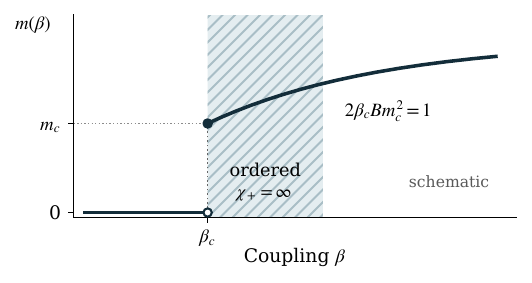}
\caption{Schematic magnetization curve. The jump at $\beta_c$
satisfies $2\beta_cBm_c^2=1$. Within the shaded ordered interval,
$\chi^+=\infty$ even after subtracting the magnetization plateau.
Such an interval exists under \eqref{main:tail}; its width and the
curve's shape are not determined. Shading ends without implying
another transition.}\label{fig:jump}
\end{figure}

\emph{Connected fluctuations and field response.---}
At an ordered coupling put
\begin{equation}\label{main:connected}
 Q(\beta)=2\beta Bm(\beta)^2,\qquad
 G_\beta(n)=\langle\sigma_0\sigma_n\rangle_+-m(\beta)^2,
\end{equation}
where $+$ denotes the maximal plus state. We subtract the plateau
to measure fluctuations within that state. Combining disconnection
estimates~\cite{AN,IN} with the Simon--Lieb boundary test~\cite{DT},
averaged over the interval radius, gives, for
every $\epsilon>0$ and sufficiently large $R$,
\begin{equation}\label{main:integrated}
 \chi_\beta^+(R):=\sum_{|n|\le R}G_\beta(n)
 \ge C_\epsilon R^{3-2Q(\beta)-\epsilon}.
\end{equation}
Under \eqref{main:tail}, this bound holds wherever $m>0$.
At criticality Eq.~\eqref{main:jump} makes its exponent arbitrarily
close to one:
\begin{equation}\label{main:critical-integrated}
 \chi_c^+(R)\ge C_\epsilon R^{1-\epsilon}.
\end{equation}
Thus $\chi_c^+(R)$ diverges as $R\to\infty$.
A uniform bound $G_{\beta_c}(n)\le C(1+|n|)^{-\delta}$,
for any $\delta>0$, would contradict \eqref{main:critical-integrated}
after summation. Divergence also persists on the ordered side:
right-continuity gives $Q(\beta)\to1$ as $\beta\downarrow\beta_c$,
so $Q<3/2$ in some interval to the right of $\beta_c$.
For small enough $\epsilon$, the lower bound then diverges.
We have obtained a sufficient condition, $Q<3/2$, rather than
a location for the endpoint of the divergence region.

For the response to a uniform field, write
$m_\beta(h)=\partial_h p_\beta(h)$ for $h>0$ and
$m_c(h)=m_{\beta_c}(h)$. Applying a field to an interval whose length
is of order $h^{-1}$, then comparing with uniform field, converts
Eq.~\eqref{main:critical-integrated} into
\begin{equation}\label{main:field}
 m_c(h)-m_c\ge c_\epsilon h^\epsilon
 \qquad\text{for every }\epsilon>0
\end{equation}
at sufficiently small $h$ (SM, Sec.~S7). Choosing $\epsilon<\delta$
rules out $m_c(h)-m_c=O(h^\delta)$ for every $\delta>0$.
Neither estimate singles out a logarithmic law.
Reflection positivity gives a pointwise correlation
bound as well for the pure inverse-square chain and the continuum model below
(SM, Sec.~S7).

\emph{From the physical magnetization to phase rigidity.---}
We explain why a critical value $Q>1$ is impossible. Fix an ordered
coupling and write $m=m(\beta)$. First delete nonnegative interactions
$D_r(n)\le J_n$ whose mass
$\ell_r=\beta\sum_nD_r(n)$ tends to zero, with the normalized mass
escaping every fixed set of separations. Then
\begin{equation}\label{main:deletion}
 p_J(0)-p_{J-D_r}(0)=m^2\ell_r+o(\ell_r).
\end{equation}
Clustering in the full plus state gives the upper bound in
Eq.~\eqref{main:deletion}. To obtain the lower bound, we first add
a positive field large enough to compensate the deleted conditional
interaction. On rescaling field by $\ell_r$, Lee--Yang analyticity extends
the plateau from these compensating fields to all positive rescaled
fields. Interpolating the couplings with GKS gives the reverse
pressure bound, again with coefficient $m^2$.

Next we compare a finite interval with the infinite chain. For the free
partition function normalized by the fair product spin law, define
\begin{equation}\label{main:boundary}
 \mathfrak b_J(N)=Np_J(0)-\log Z_J^{\mathrm f}(N).
\end{equation}
The deficit is nonnegative and nondecreasing by ferromagnetism, while
$\mathfrak b_J(N)\le\beta B\log N+o(\log N)$.
Choose $c>\beta B$. The sequence
$\mathfrak b_J(2^k)-ck\log2$ tends to $-\infty$.
At each new running minimum, every backward dyadic increment of
$\mathfrak b_J$ is bounded by the corresponding increment of
$ck\log2$. We use these record volumes for the partition-function
comparison with the depleted model. Inserting
\eqref{main:deletion} bounds the energy of the spin profile under
the original normalized free law.

To state it, let $A_N(x)=\int_0^x S_N(t)\,dt$ be the cumulative
magnetization of the spin profile rescaled to $[0,1]$.
Ferromagnetic comparison bounds every limiting derivative
by $|A'|\le m$. Write $A'=mz$, extend $z$ by zero outside $[0,1]$,
and set
\begin{equation}\label{main:smoothing}
 (P_vz)(x)=\frac v2\int_{\R}e^{-v|x-y|}z(y)\,dy.
\end{equation}
Every subsequential law along the selected volumes satisfies
\begin{equation}\label{main:energy}
 \int_0^V\left[1-\E\langle z,P_vz\rangle\right]\,dv
 \le C\log V+O(1).
\end{equation}
Here $\langle\cdot,\cdot\rangle$ is the $L^2(\R)$ pairing.
We obtain \eqref{main:energy} by taking $N\to\infty$ at fixed
frequency cutoff $V$. We can then increase $V$ within the same
subsequential law. Dividing by $V$ and using $P_vz\to z$ in $L^2$
gives $\E\|z\|_2^2=1$. Because $|z|\le1$ and $z$ is supported
on $[0,1]$, almost every limiting profile satisfies $|z|=1$
almost everywhere. Its total variation can be estimated through
\begin{equation}\label{main:variation}
 \operatorname{TV}(P_vz)
 \le2v\left[1-\langle z,P_vz\rangle\right].
\end{equation}
A logarithmic average, followed by \eqref{main:energy} and lower
semicontinuity, bounds $\E\operatorname{TV}(z)$. Here
$\operatorname{TV}$ includes the jumps to zero at the endpoints.
Thus almost surely the profile is a binary step function with
finitely many jumps; no domain structure was assumed.

An interior jump gives opposite signs on its two sides. To detect
it from a fixed cut $x$, let $E^\delta_{U,V}(x)$ require the
exponentially smoothed half-profiles to have approximately opposite
signs throughout the frequency interval $[U,V]$. For $b<B$ and
small tolerance $\delta$, the Gaussian four-functions estimate is
\begin{equation}\label{main:cut}
 \mathbb P(E^\delta_{U,V}(x))\le(V/U)^{-\alpha},\qquad
 \alpha=2\beta b m^2(1-2\delta).
\end{equation}
We construct the Gaussian sources without a spectral representation
of $J$. Their inner products are bounded by $\beta b/n^2$, which
is at most $\beta J_n$ outside finitely many separations. Those
exceptions contribute a total defect $O(V^2/N)$, removed by taking
$N\to\infty$ at fixed $U,V$. The derivation in End Matter A keeps
the full common-source norm in \eqref{main:cut}.

Assume $Q>1$, and choose $0<b<B$ and small $\delta>0$ so that
$\alpha>1$. If interior jumps have positive probability, there
is an event of positive probability with a common positive
isolation scale. Choose $U$ large enough for this scale and
$\delta$, and keep it fixed. For every sufficiently large $V$,
each profile in that event admits an interval of cuts, of length
proportional to $V^{-1}$, that pass the test. The expected measure
of successful cuts is then at least $c_0/V$ for a fixed $c_0>0$.
Yet integrating \eqref{main:cut} over $x\in[0,1]$ bounds it by
$U^\alpha V^{-\alpha}$. The two estimates are incompatible as
$V\to\infty$. Hence the limiting profiles have constant sign.
Spin-flip symmetry gives probability $1/2$ to each primitive
$A(x)=mx$ and $A(x)=-mx$ along the selected volumes.

\emph{A finite certificate closes the critical argument.---}
Assume $Q(\beta_c)>1$ and consider a large free block from the
selected sequence.
Rigidity says that its spin average is close to $m_c$ or $-m_c$,
and every one of finitely many facing-layer averages is close to
that same value. On the other hand, conditional on a
Fortuin--Kasteleyn (FK) configuration, the internal components
have independent fair spin colors. Comparing these two descriptions
through the second and fourth moments yields one component with
density close to $m_c$ throughout the block and in all the layers.
The probability of finding it tends to one. The moment identities
are given in End Matter B; only paths internal to the block enter
the definition of a component.

We call such a block good and contract its chosen component.
Between distant good blocks the coupling has an inverse-square
lower bound with coefficient approaching $\beta_cBm_c^2>1/2$.
The facing layers provide a large nearest-block coupling too.
There are still bad blocks, which must be accounted for in a
normalized Gibbs law. The dilution estimate in End Matter B does
this and yields the ordering criterion of Imbrie and Newman~\cite{IN}.
Once the criterion holds with strict inequalities at one finite
block size, continuity of the good-block probability in $\beta$
keeps it valid for a small decrease in $\beta$. We would have
order below $\beta_c$, a contradiction. Hence $Q(\beta_c)\le1$;
the established lower-jump inequality gives equality.

\emph{Continuum and quantum companion.---}
The same mechanism has a direct continuous-time form. Fix a
symmetric telegraph spin $S_t$ with flip rate $\gamma>0$ and
interaction exponent
$\frac\kappa2\iint K(t-s)S_tS_s\,dt\,ds$.
Assume
\begin{equation}\label{main:continuum}
 \begin{gathered}
 K(t)=\int_{(0,\infty)}e^{-u|t|}\mu(du),\quad
 0<\mu((0,\infty))<\infty,\\
 \int\frac{\mu(du)}u<\infty,\qquad t^2K(t)\to B>0.
 \end{gathered}
\end{equation}
For pressure defined by the source $h\int S_tdt$, the critical
magnetization obeys
\begin{equation}\label{main:continuum-jump}
 0<\kappa_c<\infty,\qquad 2\kappa_cB\,m(\kappa_c)^2=1.
\end{equation}
Atoms in $\mu$ are allowed. This is the kernel obtained by
integrating out a Gaussian bath; SM, Sec.~S6 gives the spin-boson
Hamiltonian with the corresponding coupling convention.

The continuum model obeys the bounds
\eqref{main:integrated}--\eqref{main:field}, with the correlation
sum replaced by a time integral and $Q=2\kappa Bm^2$.
It places the slow-correlation results of
Refs.~\cite{Spohn,Fontes} at the saturated critical product for
each fixed kernel. The proof is performed in continuous time,
using clipped-tail deletion, endpoint correlations for the
boundary cost, and the time lengths of FK components. In particular,
it does not pass a critical lattice limit through a continuum
limit. The quantum pressure identity is used without assuming a
ground-state vector.

The constants $C_\epsilon$ and $c_\epsilon$ in the critical bounds
may depend on $\epsilon$. A sharper description must determine
the critical decay and small-field law, and the crossover scales
needed for quantitative finite-size comparisons.

\paragraph*{Data availability.}
No experimental or numerical research data were used or produced
in this work. The Letter and Supplemental Material contain the
derivations, including the continuum verification.

\nocite{HHS,LY,GKS,FKG,Holley,AD}
\bibliography{references}

\begin{thebibliography}{19}%
\makeatletter
\providecommand \@ifxundefined [1]{%
 \@ifx{#1\undefined}
}%
\providecommand \@ifnum [1]{%
 \ifnum #1\expandafter \@firstoftwo
 \else \expandafter \@secondoftwo
 \fi
}%
\providecommand \@ifx [1]{%
 \ifx #1\expandafter \@firstoftwo
 \else \expandafter \@secondoftwo
 \fi
}%
\providecommand \natexlab [1]{#1}%
\providecommand \enquote  [1]{``#1''}%
\providecommand \bibnamefont  [1]{#1}%
\providecommand \bibfnamefont [1]{#1}%
\providecommand \citenamefont [1]{#1}%
\providecommand \href@noop [0]{\@secondoftwo}%
\providecommand \href [0]{\begingroup \@sanitize@url \@href}%
\providecommand \@href[1]{\@@startlink{#1}\@@href}%
\providecommand \@@href[1]{\endgroup#1\@@endlink}%
\providecommand \@sanitize@url [0]{\catcode `\\12\catcode `\$12\catcode
  `\&12\catcode `\#12\catcode `\^12\catcode `\_12\catcode `\%12\relax}%
\providecommand \@@startlink[1]{}%
\providecommand \@@endlink[0]{}%
\providecommand \url  [0]{\begingroup\@sanitize@url \@url }%
\providecommand \@url [1]{\endgroup\@href {#1}{\urlprefix }}%
\providecommand \urlprefix  [0]{URL }%
\providecommand \Eprint [0]{\href }%
\providecommand \doibase [0]{https://doi.org/}%
\providecommand \selectlanguage [0]{\@gobble}%
\providecommand \bibinfo  [0]{\@secondoftwo}%
\providecommand \bibfield  [0]{\@secondoftwo}%
\providecommand \translation [1]{[#1]}%
\providecommand \BibitemOpen [0]{}%
\providecommand \bibitemStop [0]{}%
\providecommand \bibitemNoStop [0]{.\EOS\space}%
\providecommand \EOS [0]{\spacefactor3000\relax}%
\providecommand \BibitemShut  [1]{\csname bibitem#1\endcsname}%
\let\auto@bib@innerbib\@empty
\bibitem [{\citenamefont {Thouless}(1969)}]{Thouless}%
  \BibitemOpen
  \bibfield  {author} {\bibinfo {author} {\bibfnamefont {D.~J.}\ \bibnamefont
  {Thouless}},\ }\bibfield  {title} {\bibinfo {title} {Long-range order in
  one-dimensional {Ising} systems},\ }\href
  {https://doi.org/10.1103/PhysRev.187.732} {\bibfield  {journal} {\bibinfo
  {journal} {Phys. Rev.}\ }\textbf {\bibinfo {volume} {187}},\ \bibinfo {pages}
  {732} (\bibinfo {year} {1969})}\BibitemShut {NoStop}%
\bibitem [{\citenamefont {Aizenman}\ \emph {et~al.}(1988)\citenamefont
  {Aizenman}, \citenamefont {Chayes}, \citenamefont {Chayes},\ and\
  \citenamefont {Newman}}]{ACCN}%
  \BibitemOpen
  \bibfield  {author} {\bibinfo {author} {\bibfnamefont {M.}~\bibnamefont
  {Aizenman}}, \bibinfo {author} {\bibfnamefont {J.~T.}\ \bibnamefont
  {Chayes}}, \bibinfo {author} {\bibfnamefont {L.}~\bibnamefont {Chayes}},\
  and\ \bibinfo {author} {\bibfnamefont {C.~M.}\ \bibnamefont {Newman}},\
  }\bibfield  {title} {\bibinfo {title} {Discontinuity of the magnetization in
  one-dimensional {$1/|x-y|^2$} {Ising} and {Potts} models},\ }\href
  {https://doi.org/10.1007/BF01022985} {\bibfield  {journal} {\bibinfo
  {journal} {J. Stat. Phys.}\ }\textbf {\bibinfo {volume} {50}},\ \bibinfo
  {pages} {1} (\bibinfo {year} {1988})}\BibitemShut {NoStop}%
\bibitem [{\citenamefont {Imbrie}\ and\ \citenamefont {Newman}(1988)}]{IN}%
  \BibitemOpen
  \bibfield  {author} {\bibinfo {author} {\bibfnamefont {J.~Z.}\ \bibnamefont
  {Imbrie}}\ and\ \bibinfo {author} {\bibfnamefont {C.~M.}\ \bibnamefont
  {Newman}},\ }\bibfield  {title} {\bibinfo {title} {An intermediate phase with
  slow decay of correlations in one dimensional {$1/|x-y|^2$} percolation,
  {Ising} and {Potts} models},\ }\href {https://doi.org/10.1007/BF01218582}
  {\bibfield  {journal} {\bibinfo  {journal} {Commun. Math. Phys.}\ }\textbf
  {\bibinfo {volume} {118}},\ \bibinfo {pages} {303} (\bibinfo {year}
  {1988})}\BibitemShut {NoStop}%
\bibitem [{\citenamefont {Anderson}\ \emph {et~al.}(1970)\citenamefont
  {Anderson}, \citenamefont {Yuval},\ and\ \citenamefont {Hamann}}]{AYH}%
  \BibitemOpen
  \bibfield  {author} {\bibinfo {author} {\bibfnamefont {P.~W.}\ \bibnamefont
  {Anderson}}, \bibinfo {author} {\bibfnamefont {G.}~\bibnamefont {Yuval}},\
  and\ \bibinfo {author} {\bibfnamefont {D.~R.}\ \bibnamefont {Hamann}},\
  }\bibfield  {title} {\bibinfo {title} {Exact results in the {Kondo} problem.
  {II}. scaling theory, qualitatively correct solution, and some new results on
  one-dimensional classical statistical models},\ }\href
  {https://doi.org/10.1103/PhysRevB.1.4464} {\bibfield  {journal} {\bibinfo
  {journal} {Phys. Rev. B}\ }\textbf {\bibinfo {volume} {1}},\ \bibinfo {pages}
  {4464} (\bibinfo {year} {1970})}\BibitemShut {NoStop}%
\bibitem [{\citenamefont {Leggett}\ \emph {et~al.}(1987)\citenamefont
  {Leggett}, \citenamefont {Chakravarty}, \citenamefont {Dorsey}, \citenamefont
  {Fisher}, \citenamefont {Garg},\ and\ \citenamefont {Zwerger}}]{Leggett}%
  \BibitemOpen
  \bibfield  {author} {\bibinfo {author} {\bibfnamefont {A.~J.}\ \bibnamefont
  {Leggett}}, \bibinfo {author} {\bibfnamefont {S.}~\bibnamefont
  {Chakravarty}}, \bibinfo {author} {\bibfnamefont {A.~T.}\ \bibnamefont
  {Dorsey}}, \bibinfo {author} {\bibfnamefont {M.~P.~A.}\ \bibnamefont
  {Fisher}}, \bibinfo {author} {\bibfnamefont {A.}~\bibnamefont {Garg}},\ and\
  \bibinfo {author} {\bibfnamefont {W.}~\bibnamefont {Zwerger}},\ }\bibfield
  {title} {\bibinfo {title} {Dynamics of the dissipative two-state system},\
  }\href {https://doi.org/10.1103/RevModPhys.59.1} {\bibfield  {journal}
  {\bibinfo  {journal} {Rev. Mod. Phys.}\ }\textbf {\bibinfo {volume} {59}},\
  \bibinfo {pages} {1} (\bibinfo {year} {1987})}\BibitemShut {NoStop}%
\bibitem [{\citenamefont {Spohn}(1989)}]{Spohn}%
  \BibitemOpen
  \bibfield  {author} {\bibinfo {author} {\bibfnamefont {H.}~\bibnamefont
  {Spohn}},\ }\bibfield  {title} {\bibinfo {title} {Ground state(s) of the
  spin-boson hamiltonian},\ }\href {https://doi.org/10.1007/BF01238859}
  {\bibfield  {journal} {\bibinfo  {journal} {Commun. Math. Phys.}\ }\textbf
  {\bibinfo {volume} {123}},\ \bibinfo {pages} {277} (\bibinfo {year}
  {1989})}\BibitemShut {NoStop}%
\bibitem [{\citenamefont {Duminil-Copin}\ \emph {et~al.}(2024)\citenamefont
  {Duminil-Copin}, \citenamefont {Garban},\ and\ \citenamefont
  {Tassion}}]{DGT}%
  \BibitemOpen
  \bibfield  {author} {\bibinfo {author} {\bibfnamefont {H.}~\bibnamefont
  {Duminil-Copin}}, \bibinfo {author} {\bibfnamefont {C.}~\bibnamefont
  {Garban}},\ and\ \bibinfo {author} {\bibfnamefont {V.}~\bibnamefont
  {Tassion}},\ }\bibfield  {title} {\bibinfo {title} {Long-range models in {1D}
  revisited},\ }\href {https://doi.org/10.1214/22-AIHP1355} {\bibfield
  {journal} {\bibinfo  {journal} {Ann. Inst. H. Poincar{\'e} Probab. Statist.}\
  }\textbf {\bibinfo {volume} {60}},\ \bibinfo {pages} {232} (\bibinfo {year}
  {2024})},\ \Eprint {https://arxiv.org/abs/2011.04642} {arXiv:2011.04642
  [math.PR]} \BibitemShut {NoStop}%
\bibitem [{\citenamefont {Luijten}\ and\ \citenamefont
  {Me{\ss}ingfeld}(2001)}]{LuijtenMessingfeld}%
  \BibitemOpen
  \bibfield  {author} {\bibinfo {author} {\bibfnamefont {E.}~\bibnamefont
  {Luijten}}\ and\ \bibinfo {author} {\bibfnamefont {H.}~\bibnamefont
  {Me{\ss}ingfeld}},\ }\bibfield  {title} {\bibinfo {title} {Criticality in one
  dimension with inverse square-law potentials},\ }\href
  {https://doi.org/10.1103/PhysRevLett.86.5305} {\bibfield  {journal} {\bibinfo
   {journal} {Phys. Rev. Lett.}\ }\textbf {\bibinfo {volume} {86}},\ \bibinfo
  {pages} {5305} (\bibinfo {year} {2001})},\ \Eprint
  {https://arxiv.org/abs/cond-mat/0104175} {arXiv:cond-mat/0104175}
  \BibitemShut {NoStop}%
\bibitem [{\citenamefont {Tower}\ \emph {et~al.}(2026)\citenamefont {Tower},
  \citenamefont {Gassama}, \citenamefont {Eick}, \citenamefont {Zhang},
  \citenamefont {Ding}, \citenamefont {Tobochnik}, \citenamefont {Pelcovits},
  \citenamefont {Kosterlitz},\ and\ \citenamefont {Ling}}]{Tower2026}%
  \BibitemOpen
  \bibfield  {author} {\bibinfo {author} {\bibfnamefont {O.}~\bibnamefont
  {Tower}}, \bibinfo {author} {\bibfnamefont {A.}~\bibnamefont {Gassama}},
  \bibinfo {author} {\bibfnamefont {J.}~\bibnamefont {Eick}}, \bibinfo {author}
  {\bibfnamefont {H.}~\bibnamefont {Zhang}}, \bibinfo {author} {\bibfnamefont
  {L.}~\bibnamefont {Ding}}, \bibinfo {author} {\bibfnamefont {J.}~\bibnamefont
  {Tobochnik}}, \bibinfo {author} {\bibfnamefont {R.~A.}\ \bibnamefont
  {Pelcovits}}, \bibinfo {author} {\bibfnamefont {J.~M.}\ \bibnamefont
  {Kosterlitz}},\ and\ \bibinfo {author} {\bibfnamefont {X.~S.}\ \bibnamefont
  {Ling}},\ }\bibfield  {title} {\bibinfo {title} {Critical temperature of the
  one-dimensional {Ising} model with inverse-square interactions},\ }\href
  {https://doi.org/10.1103/hym1-lmwt} {\bibfield  {journal} {\bibinfo
  {journal} {Phys. Rev. E}\ }\textbf {\bibinfo {volume} {114}},\ \bibinfo
  {pages} {044128} (\bibinfo {year} {2026})}\BibitemShut {NoStop}%
\bibitem [{SM()}]{SM}%
  \BibitemOpen
  \href@noop {} {}\bibinfo {note} {See Supplemental Material for
  proofs}\BibitemShut {NoStop}%
\bibitem [{\citenamefont {Aizenman}\ and\ \citenamefont {Newman}(1986)}]{AN}%
  \BibitemOpen
  \bibfield  {author} {\bibinfo {author} {\bibfnamefont {M.}~\bibnamefont
  {Aizenman}}\ and\ \bibinfo {author} {\bibfnamefont {C.~M.}\ \bibnamefont
  {Newman}},\ }\bibfield  {title} {\bibinfo {title} {Discontinuity of the
  percolation density in one dimensional {$1/|x-y|^2$} percolation models},\
  }\href {https://doi.org/10.1007/BF01205489} {\bibfield  {journal} {\bibinfo
  {journal} {Commun. Math. Phys.}\ }\textbf {\bibinfo {volume} {107}},\
  \bibinfo {pages} {611} (\bibinfo {year} {1986})}\BibitemShut {NoStop}%
\bibitem [{\citenamefont {Duminil-Copin}\ and\ \citenamefont
  {Tassion}(2016)}]{DT}%
  \BibitemOpen
  \bibfield  {author} {\bibinfo {author} {\bibfnamefont {H.}~\bibnamefont
  {Duminil-Copin}}\ and\ \bibinfo {author} {\bibfnamefont {V.}~\bibnamefont
  {Tassion}},\ }\bibfield  {title} {\bibinfo {title} {A new proof of the
  sharpness of the phase transition for {Bernoulli} percolation and the {Ising}
  model},\ }\href {https://doi.org/10.1007/s00220-015-2480-z} {\bibfield
  {journal} {\bibinfo  {journal} {Commun. Math. Phys.}\ }\textbf {\bibinfo
  {volume} {343}},\ \bibinfo {pages} {725} (\bibinfo {year} {2016})},\ \Eprint
  {https://arxiv.org/abs/1502.03050} {arXiv:1502.03050 [math.PR]} \BibitemShut
  {NoStop}%
\bibitem [{\citenamefont {Fontes}(1993)}]{Fontes}%
  \BibitemOpen
  \bibfield  {author} {\bibinfo {author} {\bibfnamefont {L.~R.~G.}\
  \bibnamefont {Fontes}},\ }\bibfield  {title} {\bibinfo {title} {An ordered
  phase with slow decay of correlations in continuum {$1/r^2$} {Ising}
  models},\ }\href {https://doi.org/10.1214/aop/1176989123} {\bibfield
  {journal} {\bibinfo  {journal} {Ann. Probab.}\ }\textbf {\bibinfo {volume}
  {21}},\ \bibinfo {pages} {1394} (\bibinfo {year} {1993})}\BibitemShut
  {NoStop}%
\bibitem [{\citenamefont {Hasler}\ \emph {et~al.}(2022)\citenamefont {Hasler},
  \citenamefont {Hinrichs},\ and\ \citenamefont {Siebert}}]{HHS}%
  \BibitemOpen
  \bibfield  {author} {\bibinfo {author} {\bibfnamefont {D.}~\bibnamefont
  {Hasler}}, \bibinfo {author} {\bibfnamefont {B.}~\bibnamefont {Hinrichs}},\
  and\ \bibinfo {author} {\bibfnamefont {O.}~\bibnamefont {Siebert}},\
  }\bibfield  {title} {\bibinfo {title} {{FKN} formula and ground state energy
  for the spin boson model with external magnetic field},\ }\href
  {https://doi.org/10.1007/s00023-022-01160-6} {\bibfield  {journal} {\bibinfo
  {journal} {Ann. Henri Poincar{\'e}}\ }\textbf {\bibinfo {volume} {23}},\
  \bibinfo {pages} {2819} (\bibinfo {year} {2022})},\ \Eprint
  {https://arxiv.org/abs/2106.08659} {arXiv:2106.08659 [math-ph]} \BibitemShut
  {NoStop}%
\bibitem [{\citenamefont {Lee}\ and\ \citenamefont {Yang}(1952)}]{LY}%
  \BibitemOpen
  \bibfield  {author} {\bibinfo {author} {\bibfnamefont {T.~D.}\ \bibnamefont
  {Lee}}\ and\ \bibinfo {author} {\bibfnamefont {C.~N.}\ \bibnamefont {Yang}},\
  }\bibfield  {title} {\bibinfo {title} {Statistical theory of equations of
  state and phase transitions. {II}. lattice gas and {Ising} model},\ }\href
  {https://doi.org/10.1103/PhysRev.87.410} {\bibfield  {journal} {\bibinfo
  {journal} {Phys. Rev.}\ }\textbf {\bibinfo {volume} {87}},\ \bibinfo {pages}
  {410} (\bibinfo {year} {1952})}\BibitemShut {NoStop}%
\bibitem [{\citenamefont {Kelly}\ and\ \citenamefont {Sherman}(1968)}]{GKS}%
  \BibitemOpen
  \bibfield  {author} {\bibinfo {author} {\bibfnamefont {D.~G.}\ \bibnamefont
  {Kelly}}\ and\ \bibinfo {author} {\bibfnamefont {S.}~\bibnamefont
  {Sherman}},\ }\bibfield  {title} {\bibinfo {title} {General {Griffiths}'
  inequalities on correlations in {Ising} ferromagnets},\ }\href
  {https://doi.org/10.1063/1.1664600} {\bibfield  {journal} {\bibinfo
  {journal} {J. Math. Phys.}\ }\textbf {\bibinfo {volume} {9}},\ \bibinfo
  {pages} {466} (\bibinfo {year} {1968})}\BibitemShut {NoStop}%
\bibitem [{\citenamefont {Fortuin}\ \emph {et~al.}(1971)\citenamefont
  {Fortuin}, \citenamefont {Kasteleyn},\ and\ \citenamefont {Ginibre}}]{FKG}%
  \BibitemOpen
  \bibfield  {author} {\bibinfo {author} {\bibfnamefont {C.~M.}\ \bibnamefont
  {Fortuin}}, \bibinfo {author} {\bibfnamefont {P.~W.}\ \bibnamefont
  {Kasteleyn}},\ and\ \bibinfo {author} {\bibfnamefont {J.}~\bibnamefont
  {Ginibre}},\ }\bibfield  {title} {\bibinfo {title} {Correlation inequalities
  on some partially ordered sets},\ }\href {https://doi.org/10.1007/BF01651330}
  {\bibfield  {journal} {\bibinfo  {journal} {Commun. Math. Phys.}\ }\textbf
  {\bibinfo {volume} {22}},\ \bibinfo {pages} {89} (\bibinfo {year}
  {1971})}\BibitemShut {NoStop}%
\bibitem [{\citenamefont {Holley}(1974)}]{Holley}%
  \BibitemOpen
  \bibfield  {author} {\bibinfo {author} {\bibfnamefont {R.}~\bibnamefont
  {Holley}},\ }\bibfield  {title} {\bibinfo {title} {Remarks on the {FKG}
  inequalities},\ }\href {https://doi.org/10.1007/BF01645980} {\bibfield
  {journal} {\bibinfo  {journal} {Commun. Math. Phys.}\ }\textbf {\bibinfo
  {volume} {36}},\ \bibinfo {pages} {227} (\bibinfo {year} {1974})}\BibitemShut
  {NoStop}%
\bibitem [{\citenamefont {Ahlswede}\ and\ \citenamefont {Daykin}(1978)}]{AD}%
  \BibitemOpen
  \bibfield  {author} {\bibinfo {author} {\bibfnamefont {R.}~\bibnamefont
  {Ahlswede}}\ and\ \bibinfo {author} {\bibfnamefont {D.~E.}\ \bibnamefont
  {Daykin}},\ }\bibfield  {title} {\bibinfo {title} {An inequality for the
  weights of two families of sets, their unions and intersections},\ }\href
  {https://doi.org/10.1007/BF00536201} {\bibfield  {journal} {\bibinfo
  {journal} {Z. Wahrscheinlichkeitstheor. Verw. Geb.}\ }\textbf {\bibinfo
  {volume} {43}},\ \bibinfo {pages} {183} (\bibinfo {year} {1978})}\BibitemShut
  {NoStop}%
\end{thebibliography}%

\onecolumngrid
\twocolumngrid
\appendix
\setcounter{secnumdepth}{1}
\setcounter{section}{0}
\renewcommand{\thesection}{\Alph{section}}
\renewcommand{\theequation}{\thesection\arabic{equation}}
\makeatletter\@addtoreset{equation}{section}\makeatother
\section*{End Matter}
\section{The sharp Gaussian interface estimate}

The estimate in Eq.~\eqref{main:cut} depends on keeping track of the
Gaussian weight. Let $\mu$ be a finite spin law invariant under spin
reversal, and let $a,d$ be odd maps into a real Hilbert space. Choose
a unit vector $\varphi$. We take $y$ to be a standard Gaussian in
the finite-dimensional span of $\varphi$ and the source values, and set
\[
 w_y=\mu\exp\{\langle y,a+d\rangle-\tfrac12\|a+d\|^2\}.
\]
Assume that $w_y$ satisfies the FKG lattice condition with a
multiplicative defect of at most $e^{4\varepsilon}$, and that the
projections of $a$ and $d$ onto $\varphi$ are increasing. The
four-functions inequality~\cite{AD}, followed by Gaussian integration,
then gives, for $t>0$,
\begin{equation}\label{em:gaussian}
 \mu\{\langle a,\varphi\rangle\ge t,
       \langle d,\varphi\rangle\le-t,\ \|a+d\|\le R\}
 \le\tfrac12 e^{2\varepsilon-2t^2+R^2/2}.
\end{equation}
Let $A$ be the event in Eq.~\eqref{em:gaussian} and put
$C=\{\langle a,\varphi\rangle\ge t,\langle d,\varphi\rangle\ge t\}$.
Four functions and Gaussian integration give
$\mu(\ind_Ae^{-\|a+d\|^2/2})\le
 e^{2\varepsilon-2t^2}\sqrt{\mu(C)\mu(-C)}$.
For this calculation, use opposite tilts
$e^{\mp2t\langle\varphi,y\rangle}$ on $C$ and $-C$.
The projection of $a+d$ is at least $2t$ on $C$ and at most $-2t$
on $-C$; completing the square and Cauchy--Schwarz produce
$e^{-2t^2}$. The two events are disjoint, so the square root is at
most $1/2$. Finally, the weight on $A$ is at least $e^{-R^2/2}$,
giving Eq.~\eqref{em:gaussian} (SM, Sec.~S4).

We apply it at a cut $x$, with spins at $t_i=(i-\tfrac12)/N$ and
sources
\[
 g_i(v)=\sqrt{\beta b}\,\frac vN e^{-v|x-t_i|},\qquad b<B,
\]
in $\mathcal H=L^2([U,V],dv/v)$.
Their Gram couplings satisfy
$\langle g_i,g_j\rangle\le\beta b/|i-j|^2$ and
$\langle g_i,g_j\rangle\le\beta bV^2/(2N^2)$.
The first bound gives domination by $\beta J$ at all sufficiently
large separations. Only finitely many bond lengths remain, and the
second bound makes their total lattice defect
$\varepsilon=O(V^2/N)$. This error vanishes as $N\to\infty$ with
$U,V$ fixed.
On the opposite-phase event, put $\varrho=m\sqrt{\beta b}$ and
$T=\log(V/U)$. The source bounds allow
$t=(1-\delta)\varrho\sqrt T$ and
$R=2\delta\varrho\sqrt T$ in Eq.~\eqref{em:gaussian}.
After adding the two orientations and taking the volume limit, the
bound is exactly $e^{-2\beta bm^2(1-2\delta)T}$.

\section{The finite ordering certificate}

For the internal FK components $C$ of a block of length $L$, let
$a_C=|C|/L$ and $b_{C,j}=|C\cap I_j|/|I_j|$ for the finitely many
facing layers $I_j$. Conditional on the components, their spin
colors are independent fair signs. Write
$A_2=\sum_Ca_C^2$, $A_4=\sum_Ca_C^4$, and let $M$ and $M_j$ be
the spin averages over the whole block and layer $I_j$, respectively.
The color moments are
\begin{equation}\label{em:colors}
 \begin{split}
 \E[(M^2-m^2)^2\mid\omega]
   &=(A_2-m^2)^2+2(A_2^2-A_4),\\
 \E[(M_j-M)^2\mid\omega]&=\sum_C(b_{C,j}-a_C)^2.
 \end{split}
\end{equation}
Primitive rigidity makes the unconditional expectations on the
left tend to zero. Since $A_2^2-A_4$ is the sum of the cross terms
$a_C^2a_{C'}^2$ with $C\ne C'$, the first line forces
$\max_C a_C\to m$ in probability.
Choose a component attaining this maximum. Its contribution
$(b_{C,j}-a_C)^2$ to the second line must tend to zero in probability,
for each $j$. All its layer densities therefore approach $m$ together
with its block density. For any $\theta<m$, it meets every quota with
probability tending to one along the selected volumes.

Fix $a_0>1/2$ below the resulting tail strength. The facing layers
provide a comparison chain with a sufficiently large $D_1$ and
$D_d=a_0/(d+1)^2$ for $d\ge2$. Before choosing any component, we
apply domination of the internal marginal to place the connectivity
lower bound under the product of the internal free laws. Under that
product law, the indicators $\lambda_i$ of good blocks are independent
with mean $p$.
Extend the site factors to $[0,1]$, and let $F(\lambda)$ be the
normalized correlation of the finite chain with couplings
$D_{ij}\lambda_i\lambda_j$. Differentiation in any one coordinate,
with the others fixed, gives by GKS
\[
 F'\ge0,\qquad F''\ge-gF',\qquad g=4\sum_{d\ge1}D_d.
\]
These inequalities make $F$ separately convex after the substitution
$\lambda_i=\eta(t_i)$, where
\begin{equation}\label{em:dilution}
 \eta(t)=-g^{-1}\log[1-(1-e^{-g})t].
\end{equation}
Applying Jensen's inequality one coordinate at a time yields
$\E F(\lambda)\ge F(\eta(p),\ldots,\eta(p))$.
When $p$ is sufficiently close to one, Ref.~\cite{IN} gives the
required finite free-volume ordering bound for $\eta(p)^2D$.
A vacant block has an independent centered spin, so its correlations
with other blocks vanish without an additional occupation factor. Summing the resulting
correlations gives a positive whole-volume second moment; the pressure
ceiling then implies positive spontaneous magnetization. We fix the
block and every comparison parameter before decreasing the coupling.
The constants and the finite-volume continuity argument are given
in SM, Sec.~S5.

\end{document}


\begin{center}
{\Large\bfseries Supplemental Material}\par
\vspace{5pt}
{\large Universal Critical Jump for Inverse-Square Ising Ferromagnets}\par
\vspace{7pt}
Dyuman Bhattacharya\par
{\small Department of Physics and Astronomy, University of Waterloo,\\
Waterloo, Ontario, Canada}\par
{\small\href{mailto:d7bhatta@uwaterloo.ca}{d7bhatta@uwaterloo.ca}}
\end{center}
\medskip
\begingroup\footnotesize\setlength{\parskip}{0pt}\tableofcontents\endgroup

\section{Models and main results}\label{master:setup}

\subsection{The unrestricted lattice theorem}

Let $\sigma_i\in\{-1,1\}$, $i\in\mathbb Z$, and assume
\begin{equation}\label{master:tail}
 J_n\in[0,\infty),\qquad n^2J_n\longrightarrow B\in(0,\infty).
\end{equation}
Every coupling is finite, so $\sum_nJ_n<\infty$. On
$\Lambda_N=\{1,\ldots,N\}$ use the normalized free partition function
\begin{equation}\label{master:partition}
 Z_{J,\beta}(N,h)=2^{-N}\sum_{\sigma\in\{-1,1\}^N}
 \exp\left\{\beta\sum_{1\le i<j\le N}J_{j-i}\sigma_i\sigma_j
                    +h\sum_{i=1}^N\sigma_i\right\}.
\end{equation}
Each unordered pair is counted once; $h$ is dimensionless. The
normalized spin law is $\mu^{\mathrm f}_{J,\beta,N,h}$, and
$\mu^+_{J,\beta,h}$ is the infinite-volume limit with exterior
spins fixed to $+1$. Define
\begin{equation}\label{master:critical-def}
 p_{J,\beta}(h)=\lim_{N\to\infty}\frac1N\log Z_{J,\beta}(N,h),\qquad
 m_J(\beta)=\langle\sigma_0\rangle^+_{J,\beta,0},\qquad
 \beta_c=\inf\{\beta>0:m_J(\beta)>0\}.
\end{equation}
We prove below that $m_J(\beta)=\partial_h^+p_{J,\beta}(0)$.
We omit a fixed $\beta$ from pressure and measure notation;
$K$ denotes other summable nonnegative interactions at that $\beta$.

\begin{theorem}[Unrestricted critical saturation]\label{master:main}
Under \eqref{master:tail}, $0<\beta_c<\infty$, $m_c:=m_J(\beta_c)>0$, and
\begin{equation}\label{master:equality}
 \boxed{2\beta_cB\,m_c^2=1.}
\end{equation}
\end{theorem}

The finite core may contain zeros; no monotonicity or tail convergence
rate is assumed. Define the rescaled spin primitive
\begin{equation}\label{master:primitive-def}
 A_N(x)=\int_0^x S_N(t)\,dt,\qquad
 S_N(t)=\sigma_i\quad\text{for }(i-1)/N<t\le i/N.
\end{equation}
\begin{theorem}[Rigidity under strict surplus]\label{master:rigidity}
If $m=m_J(\beta)>0$ and $2\beta Bm^2>1$, there is a sequence
$N_s\to\infty$ such that, under the original zero-field free laws,
\begin{equation}\label{main:rigid-limit}
 A_{N_s}\ \Longrightarrow\ \tfrac12\delta_{x\mapsto mx}
                      +\tfrac12\delta_{x\mapsto-mx}
 \quad\text{in }C[0,1].
\end{equation}
Consequently, for every finite collection of fixed intervals of
positive length in $[0,1]$, their spin averages converge jointly to
the same fair random sign times $m$.
\end{theorem}

\subsection{The continuum companion}

Fix a flip rate $\gamma>0$. Let $\Pi_{\gamma,T}$ be the law of a
stationary symmetric telegraph spin $S_t\in\{-1,1\}$ on $[0,T]$,
with fair initial sign and Poisson flips of rate $\gamma$.
Assume
\begin{equation}\label{sm:continuum-kernel}
 \begin{gathered}
 K(t)=\int_{(0,\infty)}e^{-u|t|}\,\mu(du),\qquad
 0<\mu((0,\infty))<\infty,\\
 \int_{(0,\infty)}\frac{\mu(du)}u<\infty,\qquad
 t^2K(t)\longrightarrow B\in(0,\infty).
 \end{gathered}
\end{equation}
The positive measure $\mu$ may have atoms. With $\kappa\ge0$, set
\begin{equation}\label{sm:continuum-partition}
 Z_{\kappa,K}(T,h)=\E_{\Pi_{\gamma,T}}
 \exp\left\{\frac\kappa2\int_0^T\!\int_0^T
 K(t-s)S_tS_s\,dt\,ds+h\int_0^TS_t\,dt\right\}.
\end{equation}
Write $p_\kappa(h)=\lim_{T\to\infty}T^{-1}\log Z_{\kappa,K}(T,h)$,
$m(\kappa)=\partial_h^+p_\kappa(0)$, and
$\kappa_c=\inf\{\kappa>0:m(\kappa)>0\}$, keeping $K$ and $\gamma$ fixed.

\begin{theorem}[Continuum critical saturation]
\label{master:continuum}\label{sm:continuum}
Under \eqref{sm:continuum-kernel}, $0<\kappa_c<\infty$,
$m_c:=m(\kappa_c)>0$, and
\begin{equation}\label{main:continuum-equality}
 \boxed{2\kappa_cB\,m_c^2=1.}
\end{equation}
At an ordered coupling with $2\kappa Bm(\kappa)^2>1$, the continuum
primitive $A_T(x)=T^{-1}\int_0^{Tx}S_t\,dt$ has the same fair
constant-phase subsequential limit as in \eqref{main:rigid-limit}.
\end{theorem}

\section{Thermodynamics and diffuse deletion}
\label{ms:foundations}

\subsection{Pressure and the plus state}

Fix $\beta>0$ and a summable ferromagnetic interaction $K$, and write
$M_K(h)=\langle\sigma_0\rangle^+_{K,h}$ and $m_K=M_K(0)$.
Absolute summability gives the Gibbs specification in the plus-volume
limit. We use GKS at nonnegative fields~\cite{GKS}, Holley
comparison~\cite{Holley}, and FKG~\cite{FKG}, which allows real fields.

\begin{lemma}[Thermodynamic facts]\label{ms:pressure}
\label{ms:plus-clustering}\label{ms:beta-right-continuity}
The free and plus pressures coincide. Their common value $p_K$ is
even and convex, with
\begin{align}
 |p_K(h)-p_{\widetilde K}(\widetilde h)|
 &\le\beta\sum_{d\ge1}|K_d-\widetilde K_d|+|h-\widetilde h|,
 \label{ms:pressure-lipschitz}\\
 p'_K(h)&=M_K(h)\quad(h>0),\qquad p'_K(0+)=m_K.
 \label{ms:magnetization-pressure}
\end{align}
The function $M_K$ extends holomorphically to $\operatorname{Re}h>0$
with positive real part, and the plus state clusters:
\begin{equation}
 \langle\sigma_0\sigma_d\rangle^+_{K,0}\longrightarrow m_K^2.
 \label{ms:clustering}
\end{equation}
For fixed $J$, $\beta\mapsto m_J(\beta)$ is nondecreasing and
right-continuous.
\end{lemma}
\begin{proof}
For two free intervals in the same real field, every cross-pair
mean under the product law is nonnegative. Jensen therefore makes
$\log Z_K(N,h)$ superadditive. The bound
$N^{-1}\log Z_K(N,h)\le\beta\sum_dK_d+\log\cosh h$
gives the pressure limit. Spin reversal, convexity, and exponent
comparison give the elementary pressure properties. The exterior
plus spins contribute at most
\begin{equation}
 2\beta\sum_{d\ge1}\min(d,N)K_d=o(N),
 \label{ms:plus-boundary-cost}
\end{equation}
so it does not change the pressure.

Lee--Yang~\cite{LY} places the free field-polynomial zeros
$\zeta_j$ on the unit circle, whence
\begin{equation}
 \frac1N\partial_h\log Z_K(N,h)
 =\frac1N\sum_{j=1}^N
       \frac{\zeta_j+e^{-2h}}{\zeta_j-e^{-2h}}.
 \label{ms:lee-yang-herglotz}
\end{equation}
These functions have positive real part and values in $[0,1]$ at
$h=1$. The Carath\'eodory bound makes them a normal family on the
right half-plane. Convexity identifies every subsequential limit
with $p'_K$ at its almost-everywhere real points of differentiability;
the identity theorem and integration therefore give an analytic
derivative on $h>0$. Its real part is strictly positive, since GKS
gives $p'_K(h)\ge\tanh h$ there.

Inside a large plus volume, the mean at a site lies between $M_K(h)$
and the central mean of any smaller plus interval around that site
contained in the large volume.
Average over the large volume, then exhaust the smaller interval:
both bounds approach $M_K(h)$, identifying the pressure derivative.
Continuity of the finite-interval upper bounds and monotonicity give
$M_K(h)\to m_K$ as $h\downarrow0$. Similarly, $m_J(\beta)$ is the
infimum of continuous nondecreasing finite-volume means, hence is
right-continuous.

The plus state is maximal among Gibbs states. Conditioning on a
nontrivial tail event and its complement gives two dominated Gibbs
states whose mixture is the plus state. Their expectations of
increasing cylinder functions must therefore coincide with its own.
Both conditional states equal the plus state, proving tail triviality.
For the exterior
$\sigma$-field $\mathcal F_R$, reverse martingale convergence gives
\[
 \sup_{|d|>R}\bigl|\langle\sigma_0\sigma_d\rangle^+_{K,0}-m_K^2\bigr|
 \le\E^+\bigl|\E^+[\sigma_0\mid\mathcal F_R]-m_K\bigr|
 \longrightarrow0,
\]
which proves clustering.
\end{proof}

\subsection{Diffuse-deletion response}
\label{ms:scalar-response}

The next argument also applies to the continuum once the pressure,
comparison, and clustering facts above have been established there.
Write $\lambda=\beta$ and $\int_+f=\sum_{d\ge1}f(d)$ on the
lattice, or $\lambda=\kappa$ and $\int_+f=\int_0^\infty f(t)\,dt$
in the continuum. All restricted integrals use the same measure.

\begin{theorem}[Diffuse-deletion response]\label{ms:deletion}
Let $0\le D_r\le J$ satisfy, as $r\downarrow0$,
\begin{equation}
 \ell_r:=\lambda\int_+D_r>0,\qquad \ell_r\to0,\qquad
 \frac{\int_{(0,M]}D_r}{\int_+D_r}\to0
 \quad\text{for every fixed }M<\infty.
 \label{ms:diffuse-deletion-assumptions}
\end{equation}
With $K_r=J-D_r$ and $m=M_J(0)$,
\begin{equation}
 p_J(0)-p_{K_r}(0)=m^2\ell_r+o(\ell_r).
 \label{ms:deletion-response}
\end{equation}
\end{theorem}
\begin{proof}
Put $a_r=2\ell_r$ and $M_r=M_{K_r}$. The missing lattice conditional
field is at most $a_r$. On a continuum mesh cell of width $\Delta$
it is at most $a_r\Delta$, compensated by the source $h\Delta$.
Holley and GKS therefore give
\begin{equation}
 m\le M_r(a_rt)\le M_J(a_rt)\longrightarrow m\qquad(t\ge1).
 \label{ms:plateau-large-t}
\end{equation}
The positive-real-part functions $M_r(a_rt)$ are a normal family,
bounded at $t=1$. Their limits equal $m$ on $t>1$, hence everywhere
on $\operatorname{Re}t>0$. Scaling $h=a_rt$ and using bounded
convergence therefore gives
\begin{equation}
 M_r(a_rt)\to m\quad(t>0),\qquad
 \int_0^{a_r}[M_J(h)-M_r(h)]\,dh=o(\ell_r).
 \label{ms:plateau-all-t}
\end{equation}
No assertion at $t=0$ is needed.

Let $\Delta p_r=p_J-p_{K_r}$. At fixed $r,h>0$, interpolation
in finite plus volumes gives $\Delta p_r(h)\ge\ell_rM_r(h)^2$:
GKS bounds every internal pair mean below by $M_r(h)^2$, and
exterior-field derivatives are nonnegative. Divide by volume and
exhaust before varying $r,h$. Hence
\begin{equation}
 \Delta p_r(0)=\Delta p_r(a_r)
       -\int_0^{a_r}(M_J-M_r)(h)\,dh
 \ge m^2\ell_r+o(\ell_r).
 \label{ms:trial-pressure}
\end{equation}
Conversely, free-volume interpolation and plus comparison give
\begin{equation}
 0\le\Delta p_r(0)\le
 \lambda\int_+D_r(t)\langle S_0S_t\rangle^+_J.
 \label{ms:deletion-upper}
\end{equation}
Here $S_t=\sigma_t$ on the lattice. Divide by $\ell_r$ and use
clustering: the normalized deletion mass escapes every bounded set.
This proves the matching upper bound. The continuum comparisons pass
through fixed-volume time-mesh limits, with sums replaced by integrals.
\end{proof}

\section{Boundary records and macroscopic phase regularity}
\label{ms:boundary-records}

\subsection{Boundary records and physical energy}

Write $p_K=p_K(0)$ and
\begin{equation}
 \mathfrak b_K(N)=Np_K-\log Z_K(N,0),\qquad \mathfrak b_K(0)=0.
 \label{ms:boundary-deficit-definition}
\end{equation}

\begin{lemma}[Boundary records]\label{ms:boundary-deficit}
\label{ms:records}
For every summable ferromagnetic $K$, $\mathfrak b_K$ is nonnegative
and nondecreasing, with
\begin{align}
 \mathfrak b_K(N)&\le\beta\sum_{d\ge1}\min(d,N)K_d,
 \label{ms:boundary-deficit-bound}\\
 \mathfrak b_J(N)&\le\beta B\log N+o(\log N).
 \label{ms:log-boundary}
\end{align}
For each $c>\beta B$, there is one sequence $N_s=2^{k_s}\to\infty$
such that
\begin{equation}
 \mathfrak b_J(N_s)-\mathfrak b_J(2^{-j}N_s)\le cj\log2
 \quad(0\le j\le k_s).
 \label{ms:record-increment}
\end{equation}
\end{lemma}
\begin{proof}
Adding a fair endpoint spin gives
\begin{equation}
 a_N:=\log Z_K(N+1)-\log Z_K(N)
 =\int_0^1\beta\sum_{d=1}^N
       K_d\langle\sigma_0\sigma_d\rangle_{N,t}\,dt.
 \label{ms:endpoint-increment}
\end{equation}
GKS makes $a_N$ nondecreasing under enlargement of the interval.
Its Ces\`aro mean tends to $p_K$, so $a_N\uparrow p_K$ and
$\mathfrak b_K(N+1)-\mathfrak b_K(N)=p_K-a_N\ge0$.
Partitioning $MN$ sites into $M$ blocks bounds the cross action by
$\beta M\sum_d\min(d,N)K_d$; division by $M$ and $M\to\infty$
give the deficit bound. The tail assumption makes this sum
$B\log N+o(\log N)$ for $K=J$.
Finally choose the strict running minima of
$\mathfrak b_J(2^k)-ck\log2\to-\infty$ and compare each minimum
with its value $j$ steps earlier.
\end{proof}

\begin{proposition}[Physical energy comparison]\label{ms:partition-comparison}
For any $0\le D\le J$, put $H=J-D$,
$\Delta p=p_J-p_H$, and
$E_D=\beta\sum_{i<l\le N}D(l-i)\sigma_i\sigma_l$.
On the same record sequence,
\begin{equation}
 \E^{\mathrm f}_{J,N_s}E_D
 \ge(1-2^{-j})N_s\Delta p-cj\log2
 \quad(0\le j\le k_s).
 \label{ms:record-energy}
\end{equation}
\end{proposition}
\begin{proof}
The exact partition ratio $F_N=\log Z_J(N)-\log Z_H(N)$ satisfies
\begin{align}
 F_N&=N\Delta p-\mathfrak b_J(N)+\mathfrak b_H(N),
 \label{ms:partition-lower-exact}\\
 0\le F_N&=-\log\E^{\mathrm f}_{J,N}e^{-E_D}
            \le\E^{\mathrm f}_{J,N}E_D.
 \label{ms:partition-upper-energy}
\end{align}
Subtract the first identity at $N$ and $n$; monotonicity of
$\mathfrak b_H$ and $F_n\ge0$ give
$F_N\ge(N-n)\Delta p-[\mathfrak b_J(N)-\mathfrak b_J(n)]$.
Take $N=N_s$, $n=2^{-j}N_s$. In the second line, GKS and Jensen
give the lower and upper bounds under the original normalized law.
\end{proof}

The same comparisons hold for real lengths $T\ge t$ and integrals
of nonnegative pair kernels. They require a monotone deficit and
logarithmic bound, but no spectral representation.

\subsection{The pressure ceiling}
\label{ms:physical-primitive}

The primitives \eqref{master:primitive-def} lie in the compact set
$\mathcal K_1=\{A\in C[0,1]:A(0)=0,\ \operatorname{Lip}(A)\le1\}$.
Section~\ref{ms:foundations} lets us bound the amplitude of these
limits. The notation $\int_I S$ and $|I|$ uses counting measure
on the lattice and Lebesgue measure in the continuum.

\begin{lemma}[Amplitude ceiling]\label{ms:amplitude-ceiling}
For a normalized finite free law, a measurable set $I$ in its volume,
and $h\ge0$,
\begin{equation}
 \log\E^{\mathrm f}e^{h\int_I S}
 \le |I|[p_K(h)-p_K(0)].
 \label{ms:block-pressure-ceiling}
\end{equation}
Every weak primitive limit is almost surely $m_K$-Lipschitz, and
\begin{equation}
 \limsup_{N\to\infty}\E^{\mathrm f}_{K,N}
       \left(N^{-1}\sum_{i=1}^N\sigma_i\right)^2\le m_K^2.
 \label{ms:second-moment-ceiling}
\end{equation}
The latter statement uses the normalized spin integral in the continuum.
For $K=J$ and $m>0$, each limit therefore has a measurable derivative
\begin{equation}
 A'=mz,\qquad |z|\le1\quad\text{a.e.}
 \label{ms:primitive-amplitude}
\end{equation}
\end{lemma}
\begin{proof}
In the law with field $t\ind_I$, raising the other fields to $t$
and passing to plus volumes bounds every mean in $I$ by $M_K(t)$.
Integration from $0$ to $h$ proves \eqref{ms:block-pressure-ceiling}.
Given $\varepsilon>0$, choose $h>0$ with
$p_K(h)-p_K(0)<h(m_K+\varepsilon/2)$. For $|I|>0$, Chernoff and spin reversal give
\begin{equation}
 \Pp^{\mathrm f}\left(\left||I|^{-1}\int_I S\right|>m_K+\varepsilon\right)
 \le2e^{-h\varepsilon|I|/2}.
 \label{ms:block-amplitude-concentration}
\end{equation}
Boundedness gives the whole-volume second-moment bound. For each
macroscopic interval with rational endpoints, pass to a weak limit
and let $\varepsilon\downarrow0$. Its primitive increment is bounded
in magnitude by $m_K$ times its length; lattice rounding costs at
most $2/N$. Countability gives these bounds simultaneously almost
surely. Continuity extends them to all intervals, proving the
Lipschitz bound; difference quotients give a measurable derivative.
\end{proof}

\subsection{Logarithmic energy and finite interfaces}
\label{mg:regularity}

Fix $m=m_J(\beta)>0$, $0<b<B$, and $c>\beta B$.
Choose $R_0$ with $J_d\ge b/d^2$ for $d>R_0$ and set
\begin{align}
 L_r(d)&=b\int_0^rue^{-ud}\,du,
 \label{ms:soft-deletion}\\
 D_r(d)&=\min\{J_d,L_r(d)\},\qquad H_r=J-D_r.
 \label{ms:clipped-deletion}
\end{align}
Since $L_r(d)\le\min\{b/d^2,br^2/2\}$, clipping affects only
$d\le R_0$, and
\begin{equation}
 \sum_dL_r(d)=b\int_0^r\frac{u}{e^u-1}\,du=br+O(r^2),
 \qquad \sum_d(L_r-D_r)(d)\le bR_0r^2/2.
 \label{ms:clip-core-cost}
\end{equation}
Each fixed finite set has $O(r^2)$ mass. Theorem~\ref{ms:deletion}
therefore gives
\begin{equation}
 \ell_r=\beta br+O(r^2),\qquad
 \Delta p_r:=p_J-p_{H_r}=\beta bm^2r+o(r).
 \label{ms:clipped-response}
\end{equation}
By GKS, replacing $D_r$ by $L_r$ only increases its physical
expected pair energy. Proposition~\ref{ms:partition-comparison} gives
\begin{equation}
 \beta\E^{\mathrm f}_{J,N_s}\sum_{i<l\le N_s}L_r(l-i)\sigma_i\sigma_l
 \ge(1-2^{-j})N_s\Delta p_r-cj\log2.
 \label{ms:physical-energy-budget}
\end{equation}

Take any weak primitive limit $A$ along these records, write $A'=mz$,
and extend $z$ by zero outside $[0,1]$. Define
\begin{equation}
 (P_vf)(x)=\frac v2\int_\R e^{-v|x-y|}f(y)\,dy,
 \qquad \mathcal D(v)=1-\E\langle z,P_vz\rangle\ge0.
 \label{mg:PQ}
\end{equation}
The inequality follows because $P_v$ is an $L^2$ contraction and
$\|z\|_2\le1$.
With $t_i=(i-\tfrac12)/N$ and
$\nu_N=N^{-1}\sum_i\sigma_i\delta_{t_i}$, the exact energy identity is
\begin{equation}
 \beta\sum_{i<k}L_{V/N}(k-i)\sigma_i\sigma_k
 =\frac{\beta b}{2}\int_0^V v
       \left[\iint e^{-v|x-y|}\,d\nu_N(x)d\nu_N(y)-\frac1N\right]dv.
 \label{mg:atomic-energy}
\end{equation}
With $A'$ extended by zero, put
$\Theta_V(A)=\beta b\int_0^V\langle A',P_vA'\rangle\,dv$.
The diagonal correction and spreading each atom over its cell give
\begin{equation}
 \left|\beta\sum_{i<k}L_{V/N}(k-i)\sigma_i\sigma_k-\Theta_V(A_N)\right|
 \le\frac{\beta b}{N}\left(\frac{V^2}{4}+\frac{V^3}{6}\right).
 \label{mg:energy-discretization}
\end{equation}
Indeed the kernel changes by at most $v/N$ on each cell pair.
Uniform primitive convergence implies weak convergence of the
derivative measures, whose total variations are at most one.
Thus, by approximation with product test functions, $\Theta_V$ is
bounded and continuous on $\mathcal K_1$ for fixed $V$, and
\begin{equation}
 \E^{\mathrm f}_{J,N_s}\left[\beta\sum_{i<k}L_{V/N_s}(k-i)\sigma_i\sigma_k\right]
 \longrightarrow\beta bm^2\int_0^V\E\langle z,P_vz\rangle\,dv.
 \label{mg:energy-limit}
\end{equation}
In \eqref{ms:physical-energy-budget}, let $s\to\infty$ with
$r=V/N_s$ and $V,j$ fixed. The fixed limiting law satisfies
\begin{equation}
 \int_0^V\mathcal D(v)\,dv
 \le2^{-j}V+\frac{c}{\beta bm^2}j\log2.
 \label{mg:all-j-budget}
\end{equation}
Only now take $j=\lceil\log_2V\rceil$ to obtain, with
$C=c/(\beta bm^2)$,
\begin{equation}
 \int_0^V\mathcal D(v)\,dv\le1+C\log(2V)\qquad(V\ge2).
 \label{mg:log-budget-eq}
\end{equation}

\begin{lemma}[From logarithmic energy to finite interfaces]\label{mg:binary-bv}
Let $z$ be a random function supported on $[0,1]$, with $|z|\le1$,
and define $P_v,\mathcal D$ by \eqref{mg:PQ}.
If $\int_0^V\mathcal D(v)\,dv\le C\log V+O(1)$ for some finite $C$,
then almost surely $|z|=1$ a.e. on $(0,1)$ and
\begin{equation}
 \E\TV_\R(z)\le2C.
 \label{mg:binary-bv-eq}
\end{equation}
In particular, $z$ has finitely many interior sign changes.
\end{lemma}
\begin{proof}
Approximate-identity convergence gives
$\mathcal D(v)\to1-\E\|z\|_2^2$. Its Ces\`aro limit is zero
by hypothesis, forcing $|z|=1$ almost everywhere, almost surely.
For any resulting zero-extended profile $f=z$,
\begin{equation}
 1-\langle f,P_vf\rangle
 =\frac v4\iint_{\R^2}e^{-v|x-y|}|f(x)-f(y)|^2\,dx\,dy.
 \label{mg:nonlocal-identity}
\end{equation}
The derivative kernel has integral zero and absolute value
$(v^2/2)e^{-v|x-y|}$. Since $f\in\{-1,0,1\}$ and
$|f(x)-f(y)|\le|f(x)-f(y)|^2$, subtracting $f(x)$ inside
the derivative convolution yields
\begin{equation}
 \TV_\R(P_vf)\le\frac{v^2}{2}\iint e^{-v|x-y|}|f(x)-f(y)|\,dx\,dy
 \le2v[1-\langle f,P_vf\rangle].
 \label{mg:smoothed-TV}
\end{equation}
The budget says that the logarithmic average of $v\mathcal D(v)$
is at most $C+o(1)$, so some deterministic $v_k\to\infty$ satisfy
$v_k\mathcal D(v_k)\le C+1/k$.
Lower semicontinuity of total variation in $L^1$, approximate-identity
convergence, and Fatou give
\[
 \E\TV_\R(z)\le\liminf_k\E\TV_\R(P_{v_k}z)
 \le2\liminf_kv_k\mathcal D(v_k)\le2C.
\]
For $n+1$ ordered Lebesgue points in $(0,1)$ with alternating signs,
$2n\le\TV_\R(z)$. Hence $z$ agrees a.e. with a finite step function.
The variation includes the two jumps of the zero extension:
$z=\ind_{[0,1]}$ gives $\TV_\R(z)=2$ and
$1-\langle z,P_vz\rangle=(1-e^{-v})/v$.
\end{proof}

\section{Gaussian interface bound and phase rigidity}
\label{mg:gaussian}

\subsection{A weighted Gaussian bound}

Write $s\vee u,s\wedge u$ for coordinatewise maximum and minimum.
We use the following form of the Ahlswede--Daykin theorem \cite{AD}.
\begin{lemma}[Four functions]\label{mg:four-functions}
If nonnegative functions on a finite spin lattice satisfy
$f_1(s)f_2(u)\le f_3(s\vee u)f_4(s\wedge u)$, then
\[
 \left(\sum_s f_1(s)\right)\left(\sum_s f_2(s)\right)
 \le\left(\sum_s f_3(s)\right)\left(\sum_s f_4(s)\right).
\]
\end{lemma}

\begin{lemma}[Spin-flip Gaussian bound]\label{mg:weighted-source}
Let $\mu$ be a spin-flip invariant probability on a finite spin
lattice, and let $a,d$ be odd Hilbert-space-valued maps. Put $h=a+d$.
For a unit vector $\varphi$, suppose that
$X_-=\langle\varphi,a\rangle$ and $X_+=\langle\varphi,d\rangle$
are increasing. In the Euclidean span of the source values and
$\varphi$, assume that
$w_y(s)=\mu(s)e^{\langle y,h(s)\rangle-\|h(s)\|^2/2}$ obeys
\begin{equation}
 w_y(s)w_y(u)\le e^{4\varepsilon}
 w_y(s\vee u)w_y(s\wedge u)
 \quad\text{for all }y,s,u.
 \label{mg:approximate-lattice}
\end{equation}
For $t>0$ and $A\subset\{X_-\ge t,\ X_+\le-t\}$,
\begin{equation}
 \mathbb E_\mu[\mathbf1_Ae^{-\|h\|^2/2}]
 \le\tfrac12e^{2\varepsilon-2t^2}.
 \label{mg:weighted-source-eq}
\end{equation}
In particular, if $\|h\|\le R$ on $A$, then
\begin{equation}
 \mu(A)\le\tfrac12e^{2\varepsilon-2t^2+R^2/2}.
 \label{mg:source-probability}
\end{equation}
\end{lemma}
\begin{proof}
Set $B=-A$, $C=\{X_-\ge t,X_+\ge t\}$, $D=-C$, and
$p_E(y)=\sum_{s\in E}w_y(s)$. For $s\in A,u\in B$, monotonicity
puts $s\vee u\in C$ and $s\wedge u\in D$. Spin-flip symmetry,
Cauchy--Schwarz, and Lemma~\ref{mg:four-functions} therefore give
\begin{align}
 q_A:=\mathbb E_\mu[\mathbf1_Ae^{-\|h\|^2/2}]
 &\le\sqrt{p_A(y)p_B(y)},\label{mg:flip-lower}\\
 &\le e^{2\varepsilon}\sqrt{p_C(y)p_D(y)}.
 \label{mg:pointwise-four-functions}
\end{align}
For standard Gaussian measure $\gamma$ on the stated span,
completion of the square gives
\[
 \int p_C(y)e^{-2t\langle\varphi,y\rangle}\,d\gamma(y)
 =e^{2t^2}\mathbb E_\mu[
   \mathbf1_Ce^{-2t\langle\varphi,h\rangle}]
 \le e^{-2t^2}\mu(C),
\]
and the analogous bound with $D$ and the opposite tilt. Another
Cauchy--Schwarz inequality yields
$\int\sqrt{p_Cp_D}\,d\gamma\le
e^{-2t^2}\sqrt{\mu(C)\mu(D)}\le\tfrac12e^{-2t^2}$,
since $C,D$ are disjoint spin-flips. Integrating these bounds
proves \eqref{mg:weighted-source-eq}; the norm bound proves
\eqref{mg:source-probability}, under the original probability $\mu$.
\end{proof}

\subsection{The physical deterministic-cut estimate}

For a primitive limit $A'=mz$, extend $z$ by zero and define
\begin{equation}
 L_vz(x)=v\int_0^\infty e^{-vs}z(x-s)\,ds,\qquad
 R_vz(x)=v\int_0^\infty e^{-vs}z(x+s)\,ds.
 \label{mg:half-averages}
\end{equation}
For a deterministic $x\in(0,1)$, a band $0<U<V<\infty$, and
$0<\delta<1/2$, let $E_{U,V}^{\delta}(x)$ be the event that, for
some $\eta\in\{-1,1\}$,
\begin{equation}
 \sup_{v\in[U,V]}|L_vz(x)-\eta|<\delta,\qquad
 \sup_{v\in[U,V]}|R_vz(x)+\eta|<\delta.
 \label{mg:face-event}
\end{equation}

\begin{proposition}[Physical deterministic-cut estimate]
\label{mg:limiting-face}
For every $0<b<B$, every subsequential primitive limit satisfies
\begin{equation}
 \mathbb P(E_{U,V}^{\delta}(x))\le(V/U)^{-\alpha},\qquad
 \alpha=2\beta bm^2(1-2\delta).
 \label{mg:limiting-face-eq}
\end{equation}
\end{proposition}
\begin{proof}
In $\mathcal H=L^2([U,V],dv/v)$, put
\[
 t_i=(i-1/2)/N,\qquad
 g_i(v)=\sqrt{\beta b}\,(v/N)e^{-v|x-t_i|},\qquad
 \varphi(v)=1/\sqrt{\log(V/U)},
\]
and use the odd sources
\begin{equation}
 a_{N,x}=\sum_{t_i<x}g_i\sigma_i,\qquad
 d_{N,x}=\sum_{t_i\ge x}g_i\sigma_i.
 \label{mg:physical-sources}
\end{equation}
Their projections onto $\varphi$ are increasing. For
$\mu=\mu_{J,N}^{\mathrm f}$, the weights in
Lemma~\ref{mg:weighted-source} have pair coefficients
$\beta J_{k-i}-q_{ik}$, where
\begin{equation}
 \begin{aligned}
 q_{ik}=\langle g_i,g_k\rangle
 &=\beta b\int_{U/N}^{V/N}
 u e^{-u(|i-1/2-Nx|+|k-1/2-Nx|)}\,du\\
 &\le\beta b\min\{|i-k|^{-2},V^2/(2N^2)\},\qquad i\ne k.
 \end{aligned}
 \label{mg:conditional-pair}
\end{equation}
Choose $R_0$ with $J_d\ge b/d^2$ for $d>R_0$. The total negative
pair coefficient is at most
\begin{equation}
 \sum_{i<k}(q_{ik}-\beta J_{k-i})_+
 \le\frac{\beta bR_0V^2}{2N}=: \varepsilon_{N,V}.
 \label{mg:repair-cost}
\end{equation}
For pair weights, the logarithmic lattice difference is four times
the sum of coefficients joining opposite disagreement sets; fields
and constants cancel. Thus \eqref{mg:approximate-lattice} holds.

Write $\varrho=m\sqrt{\beta b}$ and $T=\log(V/U)$.
If $a_{N,x}$ and $d_{N,x}$ are uniformly within $\delta\varrho$
of $\varrho$ and $-\varrho$, respectively, the Gaussian lemma applies
with $t=(1-\delta)\varrho\sqrt T$ and
$R=2\delta\varrho\sqrt T$. The full common-source norm gives
$-2t^2+R^2/2=-2\varrho^2(1-2\delta)T$.
Adding the two orientations yields
\begin{equation}
 \mathbb P_{J,N}^{\mathrm f}(\text{either orientation})
 \le e^{2\varepsilon_{N,V}}(V/U)^{-2\beta bm^2(1-2\delta)}.
 \label{mg:finite-face}
\end{equation}

Finally, $mL_vz(x)=\int_0^xve^{-v(x-y)}dA(y)$, with an analogous
formula for $R_v$. Integration by parts makes these maps jointly
continuous in $(A,x,v)$ in the uniform primitive topology. Midpoint
quadrature, including the cell crossed by $x$, bounds each source's
error against its corresponding $A_N$ integral by
$\sqrt{\beta b}(V^2/2+V)/N$, uniformly in the spin word and cut.
The event \eqref{mg:face-event} is open; for any fixed
$\delta'>\delta$, its $A_N$ version implies the discrete event
with tolerance $\delta'$ for all large $N$. Portmanteau and
\eqref{mg:finite-face}, first at fixed band and $\delta'<1/2$ as
$N\to\infty$, then as $\delta'\downarrow\delta$, prove the claim.
\end{proof}

\subsection{Exclusion of interior jumps}
\label{mg:jump-exclusion}

Assume $2\beta Bm^2>1$ and choose $0<b<B$, $0<\delta<1/2$ with
\begin{equation}
 \alpha=2\beta bm^2(1-2\delta)>1.
 \label{mg:sharp-exponent}
\end{equation}
\begin{proposition}[No interior phase jump]\label{mg:no-jump}
Every subsequential primitive limit along the record sequence is
almost surely constant in phase: $z=1$ or $z=-1$ a.e. on $(0,1)$.
\end{proposition}
\begin{proof}
By Lemma~\ref{mg:binary-bv}, $z$ is a binary step function
with finitely many jumps. Suppose interior jumps occur with positive
probability. Each has some dyadic isolation scale, so for at least
one fixed dyadic $\ell>0$, the following event has probability $p>0$:
some $x_0\in[4\ell,1-4\ell]$ and $\eta=\pm1$ satisfy
\[
 A(x_0+t)=A(x_0)-\eta m|t|\qquad(|t|\le4\ell).
\]
This event is closed, since the displayed identity passes to limits
along a convergent subsequence of the compact witness set
$[4\ell,1-4\ell]\times\{-1,1\}$.

Choose $a>0$ small, then fix $U$ with
$2a+2e^{-2U\ell}<\delta$. For large $V$, every cut
$|x-x_0|\le a/V$ satisfies \eqref{mg:face-event}: with the zero
extension included, each half-average has error throughout $[U,V]$ at most
\[
 2v|x-x_0|+2e^{-v(4\ell-|x-x_0|)}
 \le2a+2e^{-2U\ell}<\delta.
\]
The interval of admissible cuts has length $2a/V$.
By joint continuity we may apply Tonelli; with
\eqref{mg:limiting-face-eq}, this gives
\begin{equation}
 \frac{2ap}{V}
 \le\mathbb E\bigl|\{x:E_{U,V}^{\delta}(x)\}\bigr|
 =\int_0^1\mathbb P(E_{U,V}^{\delta}(x))\,dx
 \le U^\alpha V^{-\alpha}.
 \label{mg:jump-lower}
\end{equation}
With $U$ fixed, $p>0$ and $\alpha>1$ contradict this as $V\to\infty$.
\end{proof}

\begin{proof}[Proof of Theorem~\ref{master:rigidity}]
Take a subsequential limit law $\nu$ along the records. Its primitives
have the form $A(x)=\Sigma mx$ by Proposition~\ref{mg:no-jump}, and
spin reversal forces $\nu(\Sigma=1)=\nu(\Sigma=-1)=1/2$.
No other limit law is possible, so compactness gives convergence
along the selected sequence. For fixed $I=[a,b]$ with $a<b$, the
spin average is $(A_N(b)-A_N(a))/(b-a)+O(N^{-1})$.
Continuous mapping gives joint convergence; boundedness gives convergence
of all moments, including the whole-block second moment.
\end{proof}

\section{Finite FK certificate and critical closure}
\label{mf:section}

\subsection{FK comparison and its parameters}
\label{mf:fk-conventions}
On a finite interval $I$, the free random-cluster law is
\begin{equation}
 \phi_{I,\beta}^{\mathrm f}(\omega)
 =\frac{2^{k(\omega)}}{\mathcal Z_{I,\beta}^{\mathrm{FK}}}
   \prod_{e\in E(I)}p_e^{\omega_e}(1-p_e)^{1-\omega_e},
 \qquad p_{\{x,y\}}=1-e^{-2\beta J_{|x-y|}},
 \label{mf:fk-law}
\end{equation}
where $E(I)$ consists of unordered pairs and $k$ counts all
components. Given $\omega$, a spin configuration has probability
$2^{-k(\omega)}$ if it is constant on each component, and zero otherwise.
The factor $2^{k(\omega)}$ cancels in the joint law. With the spins
fixed, summing each edge separately gives
$(1-p_e)+p_e\ind_{\{\sigma_x=\sigma_y\}}$.
By
$e^{\beta J\sigma\tau}=e^{\beta J}[(1-p)+p\ind_{\{\sigma=\tau\}}]$,
normalization gives the free Ising law. Distinct components have
independent centered colors; hence
\begin{equation}
 \langle\sigma_x\sigma_y\rangle_{J,I}^{\mathrm f}
 =\phi_{I,\beta}^{\mathrm f}(x\leftrightarrow y).
 \label{mf:es}
\end{equation}
Comparing open and closed edge weights gives the conditional probability
\begin{equation}
 \phi(\omega_e=1\mid\omega_{E\setminus\{e\}})
 =\begin{cases}
 p_e,&\text{if its endpoints are already connected},\\[2pt]
 p_e/[p_e+2(1-p_e)],&\text{otherwise}.
 \end{cases}
 \label{mf:fk-conditional}
\end{equation}
These conditional probabilities give Holley comparison in the edge
parameters and boundary wiring. Taking limits includes zero parameters.

\begin{theorem}[External input: finite-free comparison]
\label{mf:in-input}
Fix $j_d\ge0$ for $d\ge2$ with $d^2j_d\to1$. For every $b_*>1$,
some finite $j_1^*>0$ and $v>0$ give
$\langle S_iS_j\rangle\ge v^2$ for all $1\le i,j\le M$ and
every odd $M\ge3$, under the free Ising law with exponent
$(b_*/2)\sum_{i<j}j_{j-i}S_iS_j$ and $j_1=j_1^*$.
\end{theorem}
This is Imbrie--Newman \cite[Theorem~1.2 and (1.24)]{IN}, including
both endpoints: their Hamiltonian has pair coefficient $j_d/2$
up to a constant, and their intervals $[-L,L]$ have odd length.

\label{mf:constants}
Fix $a_0>1/2$ and choose
\begin{equation}
 (2a_0)^{-1}<r_0<1,\qquad b_*=2r_0a_0>1.
 \label{mf:r0}
\end{equation}
Apply the theorem with $j_d=(d+1)^{-2}$ for $d\ge2$, and set
\begin{equation}
 D_1=a_0j_1^*,\qquad D_d=a_0/(d+1)^2\ (d\ge2),
 \qquad g=4\sum_{d\ge1}D_d.
 \label{mf:D}
\end{equation}
The chain with dimensionless couplings $r_0D_d$ thus has all
finite-free correlations at least $v^2$ on odd intervals. Put
\begin{equation}
 p_*:=\frac{1-e^{-g\sqrt{r_0}}}{1-e^{-g}}\in(0,1),\qquad
 q\in\N,\quad q\ge\max\{1,\lceil16D_1/a_0\rceil\}.
 \label{mf:thresholds}
\end{equation}
For $B_->0$ and $n_0\ge1$, suppose
\begin{equation}
 J_d\ge B_-/d^2\ (d\ge n_0),\qquad
 0<\theta<1,\qquad \beta B_-\theta^2\ge a_0,
 \label{mf:tail-hypothesis}
\end{equation}
and choose a block length
\begin{equation}
 L\in2^q\N,\qquad L\ge2^{q-1}n_0.
 \label{mf:L-min}
\end{equation}
This restriction holds on all sufficiently large dyadic record
volumes. For $\ell_j=L2^{-j}$, let
\begin{equation}
 E_j=\{\ell_j/2+1,\ldots,\ell_j\},\qquad
 E_j^*=\{L+1-i:i\in E_j\},\quad 0\le j<q.
 \label{mf:layers}
\end{equation}
Each family is pairwise disjoint, and
\begin{equation}
 |E_j|=|E_j^*|=\ell_j/2.
 \label{mf:layer-size}
\end{equation}
Let $G_L=G_L(\theta,q)$ require \emph{one} internal component $C$
to satisfy
\begin{equation}
 |C|\ge\theta L,\qquad
 |C\cap E_j|,\ |C\cap E_j^*|\ge\theta\ell_j/2
 \quad(0\le j<q).
 \label{mf:good-block}
\end{equation}
The two families may overlap; other components are allowed.

\subsection{The finite ordering certificate}
\label{mf:core-marginal}
Partition $\{1,\ldots,ML\}$ into blocks $I_1,\ldots,I_M$, with
$M\ge3$ odd. Write $H$ for the internal open edges, and put
$\mu_0=\bigotimes_i\phi_{I_i,\beta}^{\mathrm f}$.

\begin{lemma}[Original-marginal comparison]
\label{mf:original-comparison}
For fixed physical sites $x,y$, the function
$F_{xy}(H)=\phi_{ML,\beta}^{\mathrm f}(x\leftrightarrow y\mid H)$
is increasing, and
\begin{equation}
 \phi_{ML,\beta}^{\mathrm f}(x\leftrightarrow y)\ge\mu_0(F_{xy}).
 \label{mf:original-association}
\end{equation}
\end{lemma}
\begin{proof}
Setting interblock parameters to zero leaves $\mu_0$, which the
original internal marginal dominates. Given $H$, the remaining
edges have the normalized FK law with internal components wired.
Adding edges to $H$ adds wiring, making $F_{xy}$ increasing.
Integrating $F_{xy}$ proves \eqref{mf:original-association}.
We have compared fixed physical sites before making any choice of
component.
\end{proof}

\label{mf:contraction-dilution}
Choose one qualifying $C_i$ per good block by a fixed tie rule,
and put $C_i=\varnothing$ otherwise. At fixed $H$, delete other
interblock edges, then contract the selected components. Deletion
lowers connectivity; contraction identifies vertices already
connected internally. Parallel edges give the dimensionless coupling
\begin{equation}
 K_{ij}(H)=\beta\sum_{x\in C_i,\,y\in C_j}J_{|x-y|},
 \label{mf:contracted-coupling}
\end{equation}
since their combined parameter is $1-e^{-2K_{ij}}$. For good blocks
at distance $d\ge2$,
\begin{equation}
 K_{ij}\ge\frac{\beta B_-|C_i||C_j|}{(d+1)^2L^2}
 \ge\frac{a_0}{(d+1)^2}=D_d.
 \label{mf:far-coupling}
\end{equation}
For adjacent blocks, pair the facing layers $E_j^*,E_j$. Their
distances lie between $\ell_j$ and $2\ell_j$, so each contributes
\begin{equation}
 \frac{\beta B_-\theta^2(\ell_j/2)^2}{4\ell_j^2}
 \ge a_0/16.
 \label{mf:near-coupling}
\end{equation}
These edge pools are disjoint and together supply $D_1$.
Under $\mu_0$, the good-block indicators $\lambda_i$ are independent
Bernoulli variables with parameter
$p=\phi_{L,\beta}^{\mathrm f}(G_L)$. The contracted chain therefore
dominates couplings $\lambda_i\lambda_jD_{|i-j|}$; vacant sites
retain independent fair spins.

\begin{lemma}[Normalized dilution]\label{mf:dilution}
For $p\ge p_*$ and $i\ne j$,
\begin{equation}
 \E_\lambda\langle S_iS_j\rangle_{\lambda_x\lambda_yD_{|x-y|}}
 \ge\langle S_iS_j\rangle_{\eta(p)^2D_{|x-y|}}\ge v^2,
 \qquad
 \eta(t)=-g^{-1}\log\bigl(1-(1-e^{-g})t\bigr).
 \label{mf:dilution-bound}
\end{equation}
Every spin expectation is a normalized finite free expectation.
\end{lemma}
\begin{proof}
Allow site factors $s_x\in[0,1]$. With all but $s_i$ fixed, let
$f(s_i)=\langle S_A\rangle_{s_xs_yD_{|x-y|}}$, where
$S_A=\prod_{x\in A}S_x$, and
$X_i=S_i\sum_{j\ne i}D_{|i-j|}s_jS_j$.
Differentiating the normalized law gives
\begin{align}
 f'&=\Cov(S_A,X_i)\ge0,\label{mf:dilution-first}\\
 f''&=\Cov(S_A,X_i^2)-2\langle X_i\rangle f'\ge-gf'.
 \label{mf:dilution-second}
\end{align}
Indeed, $X_i,X_i^2$ have nonnegative spin-monomial coefficients,
so GKS applies, and $\|X_i\|_\infty\le g/2$.
Since $\eta(0)=0$, $\eta(1)=1$, and $\eta''=g(\eta')^2$,
$f\circ\eta$ is convex. Coordinatewise Jensen for the independent
Bernoulli factors gives the first inequality; GKS,
$\eta(p_*)^2=r_0$, and Theorem~\ref{mf:in-input} give the second.
\end{proof}

\begin{proposition}[Finite internal ordering certificate]
\label{mf:certificate}
Under \eqref{mf:tail-hypothesis}--\eqref{mf:L-min},
$\phi_{L,\beta}^{\mathrm f}(G_L)\ge p_*$ implies
$m_J(\beta)\ge\theta v>0$.
\end{proposition}
\begin{proof}
Write $C_{ij}(\lambda)$ for the diluted correlation. For fixed
$x\in I_i,y\in I_j$, contraction gives
$F_{xy}(H)\ge\ind_{\{x\in C_i,y\in C_j\}}C_{ij}(\lambda)$.
Lemma~\ref{mf:original-comparison}, summed over the two blocks,
and Lemma~\ref{mf:dilution} imply
\begin{align}
 \sum_{x\in I_i,y\in I_j}\langle\sigma_x\sigma_y\rangle_{J,ML}^{\mathrm f}
 &\ge\E_{\mu_0}|C_i||C_j|C_{ij}(\lambda)\notag\\
 &\ge\theta^2L^2\E_\lambda C_{ij}(\lambda)
 \ge\theta^2L^2v^2\quad(i\ne j).
 \label{mf:block-pairs}
\end{align}
There is no further occupancy factor: $C_{ij}=0$ if either site is
vacant. Summing over distinct blocks and using nonnegative
same-block correlations gives
\begin{equation}
 \E_{J,ML}^{\mathrm f}\left[(ML)^{-1}\sum_{x=1}^{ML}\sigma_x\right]^2
 \ge\theta^2v^2(1-M^{-1}).
 \label{mf:variance-positive}
\end{equation}
Let odd $M\to\infty$ and apply the second-moment ceiling
\eqref{ms:second-moment-ceiling}.
\end{proof}

\subsection{One component from spin moments}
\label{mf:common-component}
\begin{proposition}[Primitive convergence supplies the certificate]
\label{mf:primitive-to-good}
Fix $0<\theta<\theta'<m\le1$ and finite $q$. If the free primitives
along lengths satisfying \eqref{mf:L-min} converge to $A(x)=m\Sigma x$
with a fair sign $\Sigma$, then $\phi_{L,\beta}^{\mathrm f}(G_L)\to1$.
More precisely, for the whole-block average $X_0$ and the layer
averages $X_j$, put
$U_L=\E(X_0^2-m^2)^2$ and $V_L=\sum_j\E(X_j-X_0)^2$. Then
\begin{equation}
 \phi_{L,\beta}^{\mathrm f}(G_L^c)
 \le\frac{3U_L}{2\bigl(m^2-(\theta')^2\bigr)^2}
       +\frac{V_L}{(\theta'-\theta)^2}.
 \label{mf:common-bound}
\end{equation}
The bound holds for counting or Lebesgue component and layer masses.
\end{proposition}
\begin{proof}
Enumerate the layers as $F_1,\ldots,F_{2q}$. Conditional on the FK graph,
write $a_i=|C_i|/L$, $b_{ij}=|C_i\cap F_j|/|F_j|$, and
$A_2=\sum_i a_i^2$, $A_4=\sum_i a_i^4$. Independent component signs give
\begin{align}
 U&:=\E[(X_0^2-m^2)^2\mid\omega]
     =(A_2-m^2)^2+2(A_2^2-A_4),\label{mf:cluster-fourth}\\
 V&:=\sum_j\E[(X_j-X_0)^2\mid\omega]
     =\sum_{i,j}(b_{ij}-a_i)^2.\label{mf:cluster-difference}
\end{align}
For $a=\max_i a_i$, the inequalities $A_4\le a^2A_2$ and
$A_2\ge a^2$ yield
\begin{equation}
 U\ge(A_2-m^2)^2+2(A_2-a^2)^2
 \ge\tfrac23(m^2-a^2)^2.
 \label{mf:largest-component}
\end{equation}
If $a>\theta'$, a largest component meets the whole-block quota;
failure of any layer quota contributes at least
$(\theta'-\theta)^2$ to $V$. Thus
$G_L^c\subset\{a\le\theta'\}\cup\{V\ge(\theta'-\theta)^2\}$,
and Markov gives \eqref{mf:common-bound}. Finally, each $X_j$ is
a normalized primitive increment, so joint convergence to $m\Sigma$
and boundedness give $U_L,V_L\to0$. Every step uses only masses,
and hence also holds with Lebesgue mass.
\end{proof}

\subsection{Critical existence and saturation}
\label{mf:critical-existence}\label{mf:lower-input-section}
\begin{theorem}[External input: critical existence and lower jump]
\label{mf:accn-input}
Under \eqref{master:tail}, $0<\beta_c<\infty$, and at every $\beta>0$,
$m_J(\beta)=0$ or $2\beta Bm_J(\beta)^2\ge1$.
\end{theorem}
This is Aizenman--Chayes--Chayes--Newman
\cite[Section~1.1.2(c), (1.5)--(1.6)]{ACCN}: their Hamiltonian is
$-\frac12\sum_{x<y}\widetilde J_{y-x}(\sigma_x\sigma_y-1)$,
and their bound $\beta\widetilde B M^2\ge1$ becomes ours upon
taking $\widetilde J=2J$.

\begin{corollary}[Positive critical magnetization]
\label{mf:critical-positive}
We have $0<\beta_c<\infty$ and
$m_J(\beta_c)\ge(2\beta_cB)^{-1/2}>0$.
\end{corollary}
\begin{proof}
Monotonicity and the definition of $\beta_c$ give $m_J(\beta)>0$
for every $\beta>\beta_c$. Apply Theorem~\ref{mf:accn-input} there,
then let $\beta\downarrow\beta_c$ using the right continuity in
Lemma~\ref{ms:beta-right-continuity}.
\end{proof}

\label{mf:critical-endpoint}
\begin{proposition}[The strict-surplus contradiction]
\label{mf:critical-upper}
The rigidity of Theorem~\ref{master:rigidity} implies
$2\beta_cBm_J(\beta_c)^2\le1$.
\end{proposition}
\begin{proof}
Suppose $2\beta_cBm^2>1$ for $m=m_J(\beta_c)$. Choose
$B_-<B$, $0<\theta<\theta'<m$, and $a_0>1/2$ with
\begin{equation}
 a_0<\beta_cB_-\theta^2.
 \label{mf:strict-margin}
\end{equation}
Fix $n_0$ and all constants in
\eqref{mf:r0}--\eqref{mf:thresholds}. Theorem~\ref{master:rigidity}
and Proposition~\ref{mf:primitive-to-good} supply a finite dyadic
record length satisfying \eqref{mf:L-min} and
\begin{equation}
 \phi_{L,\beta_c}^{\mathrm f}(G_L)>p_*.
 \label{mf:frozen-witness}
\end{equation}
At fixed $L$ and $G_L$, finite sums and a positive partition function
make $\beta\mapsto\phi^{\mathrm f}_{L,\beta}(G_L)$ continuous,
including when some bonds vanish. Keep all comparison constants
fixed as well. Both strict inequalities persist at some
$\beta_0\in(0,\beta_c)$: $\beta_0B_-\theta^2\ge a_0$ and
$\phi_{L,\beta_0}^{\mathrm f}(G_L)>p_*$.
Proposition~\ref{mf:certificate} gives $m_J(\beta_0)\ge\theta v>0$,
a contradiction. Together with Corollary~\ref{mf:critical-positive},
this proves Theorem~\ref{master:main}.
\end{proof}

\section{Continuum critical saturation and the Ohmic model}
\label{new:continuum-proof}\label{sm:continuum-proof}

For the telegraph model
\eqref{sm:continuum-kernel}--\eqref{sm:continuum-partition}, fix
$\gamma>0$. We establish the thermodynamic inputs for any bounded
continuous nonnegative integrable kernel $J$, so they will apply
also after clipped deletion.

\subsection{Thermodynamics and quantum normalization}
\label{new:continuum-foundations}

On a mesh of width $\Delta$, the sampled telegraph law and the bath
have pair coefficients
\begin{equation}\label{new:continuum-mesh}
 a_\Delta=\tfrac12\log\coth(\gamma\Delta)>0,
 \qquad \kappa\int_{C_i\times C_j}J(t-s)\,ds\,dt\ge0,
\end{equation}
and the within-cell action is constant. Sampled paths converge almost
surely in $L^1$ on bounded intervals. Boundedness of the actions gives
convergence of the normalized laws and spin integrals, so GKS, FKG,
and Holley comparison pass to the continuum.

Realize all intervals as restrictions of one stationary telegraph path.
Differentiate at the right endpoint and translate it to zero. The
bounded finite-interval action gives, for $h\ge0$,
\begin{equation}\label{new:continuum-endpoint}
 \frac{d}{dT}\log Z_J(T,h)
 =h\langle S_0\rangle^{\mathrm f}_{[-T,0],h}
   +\kappa\int_0^T J(t)
       \langle S_0S_{-t}\rangle^{\mathrm f}_{[-T,0],h}\,dt
 =:a_J(T,h)
\end{equation}
for almost every $T$. Enlarging $[-T,0]$ leaves its reference
restriction unchanged; turning on the added nonnegative couplings
and fields increases each displayed expectation by GKS.
Thus $a_J(T,h)$ is nondecreasing and bounded by
$h+\kappa\int_0^\infty J$.
Since $\log Z_J(T,h)=\int_0^T a_J(s,h)\,ds$, its pressure exists and
equals $\lim_{T\to\infty}a_J(T,h)$; spin reversal covers $h<0$.
In particular,
\begin{equation}\label{new:continuum-deficit}
 \mathfrak b_J(T)=Tp_J(0)-\log Z_J(T,0)
 \quad\text{is nonnegative and nondecreasing}.
\end{equation}
There is no endpoint correction in \eqref{new:continuum-endpoint},
because the telegraph reference laws are projectively consistent.

The boundary estimate is
\begin{equation}\label{new:continuum-surface}
 \mathfrak b_J(T)\le W_J(T):=
 \kappa\int_0^\infty\min(T,t)J(t)\,dt,
 \qquad W_K(T)=\kappa B\log T+o(\log T).
\end{equation}
Indeed, the zero-field weighted transition kernel of a $T$-block has
both row sums equal to $Z_J(T,0)$ by spin reversal. The Markov
property therefore gives $Z_J(T,0)^n$ for the product of the internal
weights on $n$ consecutive blocks. Their omitted cross action is at
most $nW_J(T)$; divide by $n$ and let $n\to\infty$.

Free and plus pressures coincide. Pinning both endpoints changes
the logarithm of the partition function by a bounded amount: their
joint plus probability is at least
$1/4$ in both the free Gibbs law and the reference law by GKS.
The exterior action is at most $2W_J(T)=o(T)$.
The Lee--Yang argument of Lemma~\ref{ms:pressure} is uniform in the
mesh, since the magnetization per unit length lies in $[0,1]$ for
positive real field. Hence it gives $p'_J(h)=M_J(h)$ for $h>0$,
$p'_J(0+)=M_J(0)$, and the same holomorphic positive-real-part
extension. On each fixed interval, bounded local action makes the
conditional path laws bounded-density perturbations of telegraph
bridges, uniformly in the exterior spins. Local tightness and the
summable exterior interaction therefore give the Gibbs specification
in the plus limit. Maximality makes that state tail-trivial as in
Lemma~\ref{ms:plus-clustering}; therefore
\begin{equation}\label{new:continuum-cluster}
 \langle S_0S_t\rangle_+\longrightarrow m(\kappa)^2
 \quad(|t|\to\infty).
\end{equation}
The infimum of the continuous nondecreasing finite plus-volume means
also makes $m(\kappa)$ right-continuous.

For the physical Stieltjes kernel, the quantum interpretation is
\begin{equation}\label{new:continuum-H}
 H_\kappa(h)=\gamma(1-\sigma_x)+d\Gamma(u)
                  +\sigma_z\Phi(\sqrt\kappa)-h\sigma_z,
 \qquad \Phi(g)=a(g)+a^\dagger(g)
\end{equation}
on $\mathbb C^2\otimes\mathcal F(L^2(\mu))$.
Here $u>0$ and $1,u^{-1/2}\in L^2(\mu)$. Complex conjugation
fixes the coupling and commutes with $u$, so the real-Hilbert-space
construction of \cite[Lemma~2.1, Theorem~2.4, Corollary~2.6,
and Lemma~2.9]{HHS} applies, including when $\mu$ has atoms.
Their convention is $\varphi=\Phi/\sqrt2$; their shifted,
spin-rotated Hamiltonian equals $H_\kappa(h)/\gamma$ for
$\omega=u/\gamma$, $v=1$, $\lambda=\sqrt{2\kappa}/\gamma$,
and magnetic parameter $-h/\gamma$.
With $\Xi=|+x\rangle\otimes|0\rangle$, this gives
\begin{equation}\label{new:continuum-FKN}
 Z_K(T,h)=\langle\Xi,e^{-TH_\kappa(h)}\Xi\rangle,
 \qquad p_K(h)=-\inf\spec H_\kappa(h).
\end{equation}
The field covariance is $\kappa K$, yielding the action
$\frac\kappa2\iint KSS$; the pressure identity requires no
ground-state vector.

Spohn's equilibrium input~\cite[Theorems~2--3]{Spohn} gives
\begin{equation}\label{new:continuum-lower}
 0<\kappa_c<\infty,
 \qquad m(\kappa)>0\ \Longrightarrow\ 2\kappa Bm(\kappa)^2\ge1.
\end{equation}
His parameters are $\alpha=2\kappa B$, $\epsilon=\gamma$, and
$W=K/B$: his action
$-\frac\alpha8\iint W(t-s)(S_t-S_s)^2\,dt\,ds$ differs from ours
by a constant. Atoms are allowed: if $q:(0,1)\to(0,\infty)$ pushes
Lebesgue measure to $\mu/\mu((0,\infty))$, choose
$\omega=q$ and $\lambda=\sqrt{\mu((0,\infty))/B}$ in his
one-particle representation. Right continuity then gives
$m(\kappa_c)\ge(2\kappa_cB)^{-1/2}>0$.

\subsection{Clipped deletion and physical phase regularity}

Fix an ordered $\kappa$, put $m=m(\kappa)$, and choose $0<b<B$ and
$R_0$ with $K(t)\ge b/t^2$ for $t>R_0$. As on the lattice, set
\begin{equation}\label{new:continuum-deletion}
 L_r(t)=b\int_0^rue^{-ut}\,du,\qquad
 D_r=\min\{K,L_r\},\qquad K_r=K-D_r.
\end{equation}
Here $\int_0^\infty L_r=br$ exactly, and clipping is confined to
$[0,R_0]$, with
$\int_0^\infty(L_r-D_r)\le bR_0r^2/2$.
Each bounded distance interval carries $O(r^2)$ mass, while
$\ell_r:=\kappa\int_0^\infty D_r=\kappa br+O(r^2)$.
The omitted conditional field on a mesh cell is at most
$2\Delta\ell_r$, compensated by the uniform field $2\ell_r$.
All hypotheses of the common diffuse-response theorem
(Theorem~\ref{ms:deletion}) are therefore verified, giving
\begin{equation}\label{new:continuum-response}
 \Delta p_r:=p_K(0)-p_{K_r}(0)=m^2\ell_r+o(\ell_r)
                              =\kappa bm^2r+o(r).
\end{equation}
No Stieltjes representation for $K_r$ is needed.

For $c>\kappa B$, \eqref{new:continuum-surface} and
Lemma~\ref{ms:records} select $T_s=2^{k_s}\to\infty$ with
\begin{equation}\label{new:continuum-records}
 \mathfrak b_K(T_s)-\mathfrak b_K(2^{-j}T_s)\le cj\log2
 \quad(0\le j\le k_s).
\end{equation}
The integral version of Proposition~\ref{ms:partition-comparison}
applies: both deficits are monotone, the partition ratio uses the
original normalized telegraph law, and GKS permits $D_r\le L_r$.
It yields
\begin{equation}\label{new:continuum-energy-finite}
 \kappa\E^{\mathrm f}_{K,T_s}
 \int_{0<t<u<T_s}L_r(u-t)S_tS_u\,dt\,du
 \ge(1-2^{-j})T_s\Delta p_r-cj\log2.
\end{equation}

Set $A_T(x)=T^{-1}\int_0^{Tx}S_t\,dt$.
With Lebesgue measure, \eqref{ms:block-pressure-ceiling} and
Lemma~\ref{ms:amplitude-ceiling} give $|A'|\le m$ and
$\limsup_{T\to\infty}\E^{\mathrm f}[A_T(1)^2]\le m^2$.
The first bound holds almost surely in every primitive limit.
Write $A'=mz$, with $|z|\le1$, and set $z=0$ outside $[0,1]$.
The soft energy now has the exact identity
\begin{equation}\label{new:continuum-energy-exact}
 \kappa\int_{0<t<u<T}L_{V/T}(u-t)S_tS_u\,dt\,du
 =\frac{\kappa b}{2}\int_0^V v
       \iint_{[0,1]^2}e^{-v|x-y|}\,dA_T(x)dA_T(y)\,dv.
\end{equation}
The right side is $\Theta_V(A_T)$ from Section~\ref{mg:regularity},
with $\kappa$ in place of $\beta$. It is bounded and continuous in
the uniform primitive topology. No diagonal or quadrature correction
occurs in this identity.
Fix a subsequential record-limit law, then take $T_s\to\infty$
with $r=V/T_s$ and $V,j,b$ fixed in
\eqref{new:continuum-energy-finite}.
Equations \eqref{new:continuum-response}--\eqref{new:continuum-energy-exact}
give the common energy budget with coefficient $c/(\kappa bm^2)$.
The records and limiting law are independent of $b$, so only now let
$b\uparrow B$ to obtain
\begin{equation}\label{new:continuum-energy-limit}
 \int_0^V[1-\E\langle z,P_vz\rangle]\,dv
 \le2^{-j}V+\frac{c}{\kappa Bm^2}j\log2.
\end{equation}
Choosing $j=\lceil\log_2V\rceil$ under this fixed law and applying
Lemma~\ref{mg:binary-bv} proves
\begin{equation}\label{new:continuum-bv}
 |z|=1\ \text{a.e. on }(0,1),\qquad
 \E\TV_\R(z)\le\frac{2c}{\kappa Bm^2}<\infty.
\end{equation}

\subsection{Gaussian interface exclusion}

For a cut $x\in(0,1)$ and fixed $0<U<V<\infty$, take
$\mathcal H=L^2([U,V],dv/v)$ and the exact sources
\begin{equation}\label{new:continuum-sources}
 a_{T,x}(v)=\sqrt{\kappa b}\int_0^xve^{-v(x-y)}dA_T(y),\qquad
 d_{T,x}(v)=\sqrt{\kappa b}\int_x^1ve^{-v(y-x)}dA_T(y).
\end{equation}
Their projections on $\varphi(v)=1/\sqrt{\log(V/U)}$ are increasing.
The subtracted physical-time pair kernel is
\[
 q_T(s,t)=\kappa b\int_{U/T}^{V/T}
 u e^{-u(|Tx-s|+|Tx-t|)}\,du
 \le\min\!\left\{\frac{\kappa b}{|s-t|^2},
                         \frac{\kappa bV^2}{2T^2}\right\}.
\]
Its negative residual is confined to $|s-t|\le R_0$, hence
\begin{equation}\label{new:continuum-lattice-defect}
 \int_{0<s<t<T}[q_T(s,t)-\kappa K(t-s)]_+\,ds\,dt
 \le\frac{\kappa bR_0V^2}{2T}=:\varepsilon_{T,V}.
\end{equation}
Integrate the sources over the cells of \eqref{new:continuum-mesh}.
Each Gram coefficient is then the integral of $q_T$ over a cell pair.
The inequality $(\int f)_+\le\int f_+$ bounds the residual defect
by $4\varepsilon_{T,V}$; the ferromagnetic telegraph term cannot
increase it. Lemma~\ref{mg:weighted-source} applies even if a cut
crosses a cell: both source projections are increasing, and the
lemma does not require disjoint dependence. At fixed
$T,U,V$, convergence of the normalized mesh laws and source integrals
passes the bound to the telegraph law by open-event Portmanteau.

Put $\varrho=m\sqrt{\kappa b}$ and fix $0<\delta<1/2$.
On either oriented event with
$\|a_{T,x}-\eta\varrho\|_\infty<\delta\varrho$ and
$\|d_{T,x}+\eta\varrho\|_\infty<\delta\varrho$, use
$t=(1-\delta)\varrho\sqrt{\log(V/U)}$ and
$R=2\delta\varrho\sqrt{\log(V/U)}$ in the Gaussian lemma.
The two orientations have total probability at most
$e^{2\varepsilon_{T,V}}(V/U)^{-2\kappa bm^2(1-2\delta)}$,
including the full common-source norm.
The sources are continuous in $A_T$ by integration by parts.
Thus a second open-event limit, $T_s\to\infty$ at fixed band, gives
\begin{equation}\label{new:continuum-face-limit}
 \Pp(E_{U,V}^{\delta}(x))\le(V/U)^{-\alpha},
 \qquad \alpha=2\kappa bm^2(1-2\delta),
\end{equation}
for the common event \eqref{mg:face-event}.
If $2\kappa Bm^2>1$, choose $b<B$ and $\delta>0$ with $\alpha>1$.
The isolated-jump argument of Proposition~\ref{mg:no-jump} now applies
to \eqref{new:continuum-bv} and \eqref{new:continuum-face-limit}.
Compactness and spin reversal give, along the entire record sequence,
\begin{equation}\label{new:continuum-rigidity}
 A_{T_s}\Longrightarrow
 \tfrac12\delta_{\{x\mapsto mx\}}+
 \tfrac12\delta_{\{x\mapsto-mx\}}.
\end{equation}

\subsection{Poisson FK certificate and critical closure}
\label{new:continuum-FK}

The finite-interval FK law has reference cuts of rate $\gamma$,
unordered bridges of intensity $2\kappa K(t-s)\,ds\,dt$, and weight
$2^k$ for $k$ components. Fair component colors give the original
spin law: at zero bridge intensity $2^k$ doubles the cut rate to
$2\gamma$, whose fair colors flip at rate $\gamma$; integrating
bridges contributes
$\exp\{-2\kappa\int_{s<t}K(t-s)\ind_{\{S_s\ne S_t\}}dsdt\}$,
equal to the interaction weight up to a spin-independent constant.
Thus $\langle S_sS_t\rangle^{\mathrm f}=\phi^{\mathrm f}(s\leftrightarrow t)$.
There are finitely many cuts and bridges on each bounded interval,
and $2^k\le2^{N_{\rm cut}+1}$ is integrable.
The mesh reference cut probability is $\tanh(\gamma\Delta)$ and its
bridge probabilities are
$1-e^{-2\kappa\int_{C_i\times C_j}K}$.
They converge to the Poisson laws; uniformly bounded exponential cut
moments make the cluster weights uniformly integrable, justifying all
finite FK comparisons in the limit.

Use the constants $a_0,r_0,D,g,p_*,q,v$ of
\eqref{mf:r0}--\eqref{mf:thresholds}. Suppose $0<\theta<1$,
$K(t)\ge B_-/t^2$ for $t\ge t_0$, and
$\kappa B_-\theta^2\ge a_0>1/2$.
Take $L\ge2^{q-1}t_0$, put $\ell_j=L2^{-j}$,
$E_j=[\ell_j/2,\ell_j]$, $E_j^*=L-E_j$, and let $G_L$ require
one internal component with
\begin{equation}\label{new:continuum-quotas}
 |C|\ge\theta L,\qquad
 |C\cap E_j|,\ |C\cap E_j^*|\ge\theta\ell_j/2\quad(0\le j<q),
\end{equation}
using Lebesgue masses.
Severing the deterministic connections at block endpoints decreases
connectivity; deleting interblock bridges then gives the product of
internal free laws. Mesh FK monotonicity, with the integrable bound
$2^k\le2^{N_{\rm cut}+M}$ for $M$ blocks, supplies precisely the
original-marginal and conditional-wiring comparison of
Lemma~\ref{mf:original-comparison}, before component selection.
After contraction the coupling is
$\kappa\int_{C_i\times C_j}K(t-s)\,ds\,dt$.
It is at least $D_d=a_0/(d+1)^2$ for $d\ge2$; each of the $q$
disjoint adjacent facing-layer pools contributes at least $a_0/16$,
so together they supply $D_1$.
The resulting block spins are the same normalized diluted chain as
in Lemma~\ref{mf:dilution}. Integrating the fixed-point comparison
over block pairs gives \eqref{mf:variance-positive} with
$Y_M=(ML)^{-1}\int_0^{ML}S_tdt$. The common pressure ceiling proves
\begin{equation}\label{new:continuum-certificate}
 \phi^{\mathrm f}_{L,\kappa}(G_L)\ge p_*
 \quad\Longrightarrow\quad m(\kappa)\ge\theta v>0.
\end{equation}

For $0<\theta<m$, choose $\theta'\in(\theta,m)$.
The cluster-color identities and bound \eqref{mf:common-bound}
depend only on component and layer masses. With Lebesgue masses,
\eqref{new:continuum-rigidity} makes every fixed layer average tend
jointly to the same $m\Sigma$, so their bounded moments give
$\phi^{\mathrm f}_{L,\kappa}(G_L)\to1$ along the records.
For fixed $L$, couple the same cuts and common bridges at
$\kappa,\kappa'$. The unmatched bridge intensity is
$\Delta=2|\kappa-\kappa'|\int_0^L(L-t)K(t)\,dt$;
$\E2^{N_{\rm cut}+1}=2e^{\gamma L}$ gives
\begin{equation}\label{new:continuum-continuity}
 \|\phi^{\mathrm f}_{L,\kappa}-\phi^{\mathrm f}_{L,\kappa'}\|_{\rm TV}
 \le8e^{\gamma L}(1-e^{-\Delta})\longrightarrow0.
\end{equation}

If $2\kappa_cBm(\kappa_c)^2>1$, choose $0<B_-<B$, $0<\theta<m(\kappa_c)$,
and $a_0>1/2$ with $\kappa_cB_-\theta^2>a_0$.
Fix the comparison constants and a finite record block with
$\phi^{\mathrm f}_{L,\kappa_c}(G_L)>p_*$.
At fixed $L$, \eqref{new:continuum-continuity} preserves this strict
inequality for some $\kappa_0\in(0,\kappa_c)$ with
$\kappa_0B_-\theta^2>a_0$. Then \eqref{new:continuum-certificate}
gives $m(\kappa_0)>0$, a contradiction. Together with
\eqref{new:continuum-lower}, this proves Theorem~\ref{master:continuum}:
\begin{equation}\label{new:continuum-equality}
 \boxed{2\kappa_cB\,m(\kappa_c)^2=1.}
\end{equation}
We removed the time meshes with volume and parameters fixed.
The critical contradiction then required only one finite continuum block.

\section{Connected fluctuations and field response}\label{sm:response}

Fix an ordered coupling and write
\[
 \begin{array}{lll}
 m=m_J(\beta),&Q=2\beta Bm^2,&G(n)=\langle\sigma_0\sigma_n\rangle_+-m^2
       \quad\text{(lattice)},\\
 m=m(\kappa),&Q=2\kappa Bm^2,&G(t)=\langle S_0S_t\rangle_+-m^2
       \quad\text{(continuum)}.
 \end{array}
\]
Thus $m>0$ and $Q\ge1$.
Except for the pointwise extension, the lattice assumptions are only
$J_n\ge0$ and $n^2J_n\to B>0$; the continuum uses
\eqref{sm:continuum-kernel}.
Below, position integrals mean sums on the lattice, where
$S_x=\sigma_x$. Constants may depend on the fixed model and coupling.

\subsection{Disconnection and the ordering boundary test}

Let $\phi^w$ be the infinite wired FK law. For $I_L=[-L,L]$
(intersected with $\mathbb Z$ on the lattice), put
\[
 E_L=\{I_L\not\leftrightarrow\infty\},\qquad
 \tau_L(0,x)=\phi^f_{I_L}(0\leftrightarrow x),\qquad
 \tau'(0,x)=\phi^w(0\leftrightarrow x\text{ in a finite component}).
\]
FK coloring and association give
\begin{equation}\label{cs:truncated}
 G(x)=\tau'(0,x)+\Cov_{\phi^w}
   (\ind_{\{0\leftrightarrow\infty\}},\ind_{\{x\leftrightarrow\infty\}})
 \ge\tau'(0,x)\ge\phi^w(E_L)\tau_L(0,x),\qquad x\in I_L.
\end{equation}
Let $\mathcal F_L$ be the $\sigma$-field generated by the exterior
configuration and all attachments to $I_L$. Finitely many attachments are occupied.
Unless one leads to an infinite exterior component, internal edges
can join only finitely many finite components. Thus $E_L\in\mathcal F_L$.
Boundary wiring gives
$\phi^w(0\leftrightarrow x\mid\mathcal F_L)\ge\tau_L(0,x)$.
On $E_L$ this connection is finite, so multiply by its indicator and
average to obtain the last inequality~\cite[Proposition~2.1]{IN}.

\begin{lemma}[Interval disconnection]\label{cs:disconnection}
For every $\eta>0$ and all sufficiently large $L$,
\begin{equation}\label{cs:disconnection-bound}
 \phi^w(E_L)\ge c_\eta L^{-p},\qquad p=2(Q-1)+\eta.
\end{equation}
\end{lemma}
\begin{proof}
Imbrie--Newman~\cite[Proposition~2.3]{IN} applies on the lattice at
positive magnetization under our nonnegative, translation-invariant
inverse-square-tail hypothesis. Their pair coefficient is $\beta^{\rm IN}J_n^{\rm IN}/2$,
with $n^2J_n^{\rm IN}\to1$; setting
$J_n^{\rm IN}=J_n/B$, $\beta^{\rm IN}=2\beta B$ gives $2(Q-1)$.

Use the continuum FK normalization of Section~\ref{new:continuum-FK}
and write $M_H\downarrow m$ for the wired radius-$H$ connection
probability. A bridge is $H$-anchored in Spohn's terminology if
removing it leaves both endpoints connected outside their respective
radius-$H$ neighborhoods. The anchored comparison~\cite[Lemma~4.2]{AN}, with
Spohn's continuum modifications~\cite[Sections~6--7]{Spohn}, gives
\begin{equation}\label{resp:anchored-half}
 \phi^w(F^H_{L,2})\ge cL^{1-a}\qquad(a>2\kappa BM_H^2).
\end{equation}
The event in this bound requires an actual cut $\xi\in[L,2L]$
with no anchored bridge from $[0,\xi]$ to $(\xi,\infty)$.
Here the comparison applies because the FK law has strong FKG and
positive cut rate, while $K$ is bounded, decreasing, and has an
inverse-square tail; the parameters are those in
\eqref{new:continuum-lower}. The conditional bound for an interval
without cuts is exponential by Spohn's Lemma~7.1.

For the independent bridge process used in this comparison, the
renewal bound of Ref.~\cite{Fontes} has the following direct proof.
Take an independent bridge kernel
$w$ equal to $a/t^2$ beyond a bounded core and cuts of rate $\lambda>0$,
and put $I(t)=\int_0^t dx\int_t^\infty dy\,w(y-x)
=a\log(1+t)+O(1)$.
If $N$ counts cuts in $[L,2L]$ with no bridge from $[0,\xi]$ to
$(\xi,\infty)$, the Poisson formulas give
\[
 \E N=\lambda\int_L^{2L}e^{-I(x)}dx\asymp L^{1-a},\qquad
 \E N(N-1)=2\lambda^2\int_L^{2L}dx\int_x^{2L}dy\,
                  e^{-I(x)-I(y-x)}\le C\E N.
\]
The last inequality uses $a>1$, and
$\Pp(N>0)\ge(\E N)^2/\E N^2\ge cL^{1-a}$.

Choose $a>Q$ with $2(a-1)<p$, then $H$ with $2\kappa BM_H^2<a$.
Intersect $F^H_{T,2}$ with its reflection about $T/2$, and forbid
ordinary bridges from $(-T,0)$ to $(T,\infty)$ and from $(T,2T)$
to $(-\infty,0)$.
This produces $D=[\xi',\xi]\supset[0,T]$ with actual endpoint cuts
and no anchored bridge to its complement. The added exclusions cost
fixed factors, since their reference intensities are bounded, e.g.
$2\kappa\int_{-T}^0ds\int_T^\infty dt\,K(t-s)=O(1)$.
Conditional bridge intensities are at most their reference values;
FKG for these decreasing events therefore gives probability
at least $cT^{-2(a-1)}$.
Translate the construction at $T=3L$ to cover $[-L,2L]$, and also
forbid bridges from $[-H,L+H]$ to $[-L,2L]^c$, another bounded cost
when $L>2H$. A simple infinite path starting in $[0,L]$ would leave
$D$ by a bridge whose inner endpoint is outside $[-H,L+H]$.
Removing the bridge leaves one branch reaching $[0,L]$, farther
than $H$ from its inner endpoint, and another going to infinity.
It would therefore be anchored. The endpoint cuts leave no route
out along the line. Any infinite component contains a simple infinite
path, since the FK graph is locally finite. Hence $[0,L]$ misses
infinity, and translation and rescaling prove \eqref{cs:disconnection-bound}.
\end{proof}

The modified Simon--Lieb inequality~\cite[Lemma~2.7]{DT} gives the
common ordering test
\begin{equation}\label{resp:boundary-lattice}
 1\le\alpha_L:=\int_{I_L}\tau_L(0,x)\,\nu_L(dx),
\end{equation}
where
\[
 \begin{aligned}
 \nu_L(\{x\})&=\beta\sum_{y\notin I_L}J_{|x-y|}
       &&\text{(lattice)},\\
 \nu_L&=\delta_{-L}+\delta_L+
 \kappa\ind_{I_L}(x)[F(L-x)+F(L+x)]\,dx
       &&\text{(continuum)},
 \end{aligned}
\]
and $F(s)=\int_s^\infty K(t)\,dt$.
On the lattice send the remote marked spin to infinity, use plus
clustering, divide by $m^2$, and bound $\tanh(\beta J)\le\beta J$;
summability justifies dominated convergence.
For the continuum first use the mesh \eqref{new:continuum-mesh}
at fixed outer volume with a plus ghost spin. Its two telegraph
edges crossing $I_L$ contribute endpoint terms because
$\tanh a_\Delta\to1$; bath coefficients are bounded by their
couplings, using $\tanh(a+b)\le\tanh a+b$ on a shared edge.
Take the mesh limit, then exhaust the outer volume, and finally send
the marked point to infinity. Integrability and
\eqref{new:continuum-cluster} justify these limits.

The tail bounds imply
\begin{equation}\label{resp:boundary-weights}
 \nu_L(I_L)\le C(1+\log L),\qquad
 \nu_L(\{|x|<r\})\le\frac{Cr}{L-r},\quad 1\le r\le L/2.
\end{equation}
On the lattice count $2\min(n,2L+1)$ crossing pairs at distance
$n$, and sum tails beginning at $L-r$ for the central mass.
In the continuum integrate $F(s)\le C/(1+s)$; its endpoint atoms
are outside the central interval.

\subsection{Integrated fluctuations and field response}

Put $\mathcal G(R)=\int_{I_R}G(x)\,dx$ and let $m(h)$ denote the
magnetization at the fixed coupling in uniform field with source
$h\int S$.
\begin{proposition}[Integrated and field lower bounds]
\label{reset:averaged-simon-prop}\label{reset:critical-field}
For every $\epsilon>0$, at all sufficiently large $R$ and small $h>0$,
\begin{align}
 \mathcal G(R)&\ge c_\epsilon R^{3-2Q-\epsilon},
       \label{reset:integrated-G}\\
 m(h)-m(0+)&\ge c'_\epsilon h^{2(Q-1)+\epsilon}.
       \label{resp:field-bound}
\end{align}
\end{proposition}
\begin{proof}
Put $p=2(Q-1)+\eta$. By \eqref{cs:truncated}--\eqref{cs:disconnection-bound},
$\tau_L(0,x)\le C_\eta L^pG(x)$.
Average the boundary test over $R\le L<2R$, with integer $R$ on the
lattice and counting or Lebesgue measure respectively. A pair $x,y$ is separated by at most
$\min(|x-y|,R)$ radii. The interaction factor is therefore
$2\beta\sum_{d\ge1}\min(d,R)J_d$ or
$2\kappa\int_0^\infty\min(t,R)K(t)\,dt$, both $O(1+\log R)$.
The continuum endpoint terms add at most
$C_\eta R^{p-1}\mathcal G(2R)$. Thus
\begin{equation}\label{resp:averaged-boundary}
 1\le C_\eta R^{p-1}(1+\log R)\mathcal G(2R).
\end{equation}
Take $\eta=\epsilon/2$, absorb the logarithm, and rescale to prove
\eqref{reset:integrated-G}.

For the field bound set $X=\int_{I_R}S$ and
$Y=X+|I_R|\in[0,2|I_R|]$.
The normalized tilt $e^{hY}$ of the plus state is the plus-selected state with
field $h$ restricted to $I_R$; comparison with uniform field gives
\[
 m(h)-m(0+)\ge
 \frac{\Cov_+(S_0,e^{hY})}{\langle e^{hY}\rangle_+}
 \ge h e^{-2h|I_R|}\mathcal G(R).
\]
Indeed $e^{hy}-hy$ is increasing for $y\ge0$, so FKG bounds the
numerator below by $h\Cov_+(S_0,Y)$, while the denominator is at
most $e^{2h|I_R|}$. Take $R=\lfloor h^{-1}\rfloor$ and apply
\eqref{reset:integrated-G}.
\end{proof}

\begin{corollary}[Critical and adjacent ordered response]\label{resp:critical}
The connected susceptibility $\chi_+=\lim_{R\to\infty}\mathcal G(R)$
is infinite when $Q<3/2$, hence at criticality and throughout a
nonempty adjacent ordered interval.
At criticality neither $G(r)$ nor $m_c(h)-m_c$ has a uniform upper
bound by a positive power of $r^{-1}$ or $h$, respectively.
\end{corollary}
\begin{proof}
For $Q<3/2$ choose $\epsilon<3-2Q$ in
\eqref{reset:integrated-G}. Critical saturation gives $Q_c=1$,
and right continuity of $m$ gives the adjacent interval; the
sufficient condition $Q<3/2$ does not locate its endpoint.
At criticality $\mathcal G(R)\ge c_\epsilon R^{1-\epsilon}$.
An upper bound $G(r)\le C(1+|r|)^{-\delta}$ would instead give
$\mathcal G(R)\le C'[1+\int_1^R t^{-\delta}dt]
=o(R^{1-\epsilon})$ for $0<\epsilon<\min\{\delta,1\}$.
Likewise \eqref{resp:field-bound} with $\epsilon<\delta$ excludes
$m_c(h)-m_c=O(h^\delta)$.
\end{proof}

\subsection{Pointwise correlations}

\begin{lemma}[Monotonicity from reflection]\label{resp:reflection}
The connected correlation $G(r)$ is nonincreasing for nonnegative
distances in the pure inverse-square chain $J_n=B/n^2$ and in the
continuum model \eqref{sm:continuum-kernel}.
\end{lemma}
\begin{proof}
Put $c=G+m^2$, so $0<m^2\le c\le1$. The representations
\[
 J_n=B\int_0^\infty ue^{-un}\,du\quad(n\ge1),\qquad
 K(t)=\int e^{-u|t|}\,\mu(du)
\]
make each cross action in a symmetric finite plus volume a positive
integral of products of reflected half-sources, formed with weight
$e^{-ux}$ at distance $x$ from the reflection point.
For reflection through a site or continuum point, first condition on
its spin. The reference halves are then independent and identical;
for lattice bond reflection, the product reference law already factors.
Internal interactions and exterior plus fields give identical half
weights. Expanding the cross exponential for positive atomic spectral
measures expresses each reflection quadratic form as a sum of
nonnegative squares. Approximation on compact frequency intervals
and dominated convergence extend this to the displayed measures.
For the lattice, every cross separation is at least one, so the
infinite total mass of $Bu\,du$ causes no divergence. Summing over
the conditioned spin and exhausting the plus volumes preserve
reflection positivity.

On the lattice, reflection gives positive semidefinite matrices
$[c(i+j)]_{i,j\ge0}$ and $[c(i+j+1)]_{i,j\ge0}$. Their
$2\times2$ minors yield $c(n)^2\le c(n-1)c(n+1)$ for every $n\ge1$.
Thus $f=\log c$ has nondecreasing discrete slopes.
In continuous time, positivity of $[c(t_i+t_j)]$ gives
$f((s+t)/2)\le[f(s)+f(t)]/2$, and continuity makes $f$ convex.
In either case, a positive slope would force $f$ to grow without
bound at large distances, although $f\le0$.
Thus $c$, and also $G=c-m^2$, is nonincreasing.
\end{proof}

\begin{proposition}[Pointwise lower bound]\label{reset:correlation-prop}
If $G$ is nonincreasing at nonnegative distances, then for every
$\epsilon>0$ and all sufficiently large distances,
\begin{equation}\label{resp:pointwise}
 G(r)\ge c_\epsilon r^{-2(Q-1)-\epsilon}.
\end{equation}
In particular this holds for every lattice distance in the pure
inverse-square chain and every continuum distance under
\eqref{sm:continuum-kernel}.
\end{proposition}
\begin{proof}
Take $L\asymp Dr$, with fixed $D$ large enough that
\eqref{resp:boundary-weights} gives
$\nu_L(\{|x|<r\})\le1/2$. Since $\tau_L\le1$, the ordering test
\eqref{resp:boundary-lattice} leaves a contribution at least $1/2$
outside this central interval. There,
\eqref{cs:truncated}--\eqref{cs:disconnection-bound} and monotonicity
give $\tau_L(0,x)\le C_\eta L^{2(Q-1)+\eta}G(r)$.
Consequently
\[
 \tfrac12\le C_\eta L^{2(Q-1)+\eta}(1+\log L)G(r).
\]
Set $\eta=\epsilon/2$ and absorb the logarithm.
\end{proof}

At criticality, \eqref{resp:pointwise} gives
$G(r)\ge c_\epsilon r^{-\epsilon}$ for every $\epsilon>0$ in
these two models.

\bibliographystyle{apsrev4-2}
\bibliography{references}